\documentclass[12pt,preprint]{aastex}

\shorttitle{M31 and M33 Luminous Stars}
\shortauthors{Humphreys et al. }

\begin{document}

\title{Luminous and Variable Stars in M31 and M33. II. Luminous Blue Variables, Candidate LBVs, Fe II Emission Line Stars, and Other Supergiants\altaffilmark{1}}

\author{
Roberta M. Humphreys\altaffilmark{2}, 
Kerstin Weis\altaffilmark{3},
Kris Davidson\altaffilmark{2},
D.~J. Bomans\altaffilmark{3}
and
Birgitta Burggraf\altaffilmark{3} 
}

\altaffiltext{1}  
{Based  on observations  with the Multiple Mirror Telescope, a joint facility of the Smithsonian Institution and the University of Arizona and on observations obtained with the Large Binocular Telescope (LBT), an international collaboration among institutions in the United
States, Italy and Germany. LBT Corporation partners are: The University of
Arizona on behalf of the Arizona university system; Istituto Nazionale di
Astrofisica, Italy; LBT Beteiligungsgesellschaft, Germany, representing the
Max-Planck Society, the Astrophysical Institute Potsdam, and Heidelberg
University; The Ohio State University, and The Research Corporation, on
behalf of The University of Notre Dame, University of Minnesota and
University of Virginia.
} 

\altaffiltext{2}
{Minnesota Institute for Astrophysics, 116 Church St SE, University of Minnesota
, Minneapolis, MN 55455; roberta@umn.edu} 

\altaffiltext{3}
{Astronomical Institute, Ruhr-Universitaet Bochum, Germany, 
kweis@astro.rub.de}

\begin{abstract}
An increasing number of non-terminal eruptions are being found in the numerous surveys for optical 
transients. Very little is known  about these giant  eruptions, their progenitors and their
evolutionary state. A greatly improved census of the likely 
progenitor class, including the most luminous evolved 
stars, the Luminous Blue Varaibles (LBVs), and the warm and cool hypergiants is now needed 
for a complete  
picture of the final pre-SN stages of very massive stars. We have begun a survey of the evolved and unstable luminous star 
populations in several nearby resolved galaxies. In this second paper on M31 and M33, we review 
the spectral characteristics,  spectral energy distributions, circumstellar
ejecta, and evidence for mass loss for 82 luminous and variable stars. 
 We show that many of these stars have warm circumstellar dust including several of the Fe II emission line stars, but conclude that the confirmed LBVs in M31 and M33 do not.  The confirmed LBVs have 
relatively low wind speeds even in their hot, quiescent or visual minimum state compared to the 
B-type supergiants and Of/WN stars which they spectroscopically resemble. The nature of the Fe II emission line stars and their relation to the LBV state remains uncertain, but some  have 
properties in common with the warm hypergiants and the sgB[e] stars. 
Several individual stars are discussed in detail.  We identify three possible candidate LBVs and three 
additional post-red supergiant candidates. 
We suggest that M33-013406.63 (UIT301,B416)  is not an LBV/S Dor variable, but is a very luminous 
late O-type supergiant
and one of the most luminous stars or pair of stars in M33.  
\end{abstract} 


\section{Introduction -- Luminous Blue Variables, Hypergiants, and Supernova Impostors}

An increasing number of objects initially classified as SNe in current surveys are not 
true SNe, i.e. the ``SN impostors''. Some of them appear to be pre-SN extreme events or giant eruptions like $\eta$
  Car. Others may be related to the classical Luminous Blue Variables (LBVs) such as S Dor and AG Car \citep{HD94}, while those 
that are heavily obscured may be in a post-AGB or post red supergiant stage, i.e. the Intermediate Luminosity Red Transients (ILRTs, \citet{Bond}). There is considerable diversity in 
their observed properties; maximum luminosity and duration of the outburst. 
This diversity is the sign of a little explored field in stellar astrophysics. 
Very little is known about the progenitors of the giant eruptions and their evolutionary state. 

Many of the SN impostors were initially classed
as Type IIn supernovae because of their narrow hydrogen emission lines
which are usually ascribed to shock/circumstellar matter
interactions. The coincidence of hydrogen envelopes, similar kinetic
energies, and circumstellar material make the spectra of these objects 
and of Type IIn's look very similar.   Only with subsequent observations
that confirm that they are sub-luminous, or with photometry or spectroscopy
showing that they do not develop as true SNe, are they recognized as
impostors. See \citet{vandyk}  for a recent review. 

Many authors  refer to these objects as LBVs, but most of the eruptions 
do not resemble the  classical or normal LBV   maximum light stage \citep{HD94,Vink}. 
In quiescence an LBV or S Doradus variable is a moderately evolved hot star, 
with a B-type supergiant or  Of-type/late-WN classification.  
An LBV eruption causes 
the wind to become dense and opaque, some times called a pseudo-photosphere at 
$T \sim$ 7000-8000 K resembling the spectrum of an F-type supergiant.  
On an HR diagram the object thus appears to move 
toward the right.   Since this alters the bolometric correction, 
the visual brightness increases by $\sim$  2 magnitudes while the  total luminosity 
remains nearly constant (Wolf (1989), Humphreys \& Davidson (1994) and numerous early references therein) or may decrease \citep{Groh}. Such an event can last for 
several years or even decades.  Basic causes remain somewhat mysterious; most proposed explanations 
invoke an opacity-modified Eddington limit, 
subphotospheric gravity-mode instabilities, super-Eddington winds, and envelope inflation close to the 
Eddington limit (see \S {5} in Humphreys \& Davidson(1994),Glatzel(2005),Owocki\& Shaviv (2012), and Vink(2012)).

In rare cases, however, the luminosity substantially increases during outburst; 
these have been called giant eruption LBVs \citep{HD94}, $\eta$ Car variables 
\citep{RMH99}, or $\eta$ Car analogs \citep{VanDyk05}. 
The distinction between giant eruptions and the more common LBV 
or S Dor-type variability is often overlooked in the literature. They may be 
related and originate from similar types of stars, perhaps in the same 
evolutionary stage, but the physical cause may be different. Certainly the 
energetics of the eruptions and what we observe are different.
 There are numerous questions  about the origin of the 
 giant eruptions, their relation to normal LBV outbursts, and perhaps even to SNe.
These eruptions are important.  They may account for considerable mass loss and they indicate 
that some instability has been overlooked in stellar theory. 

Very little is known about the progenitors of these giant eruptions and their 
evolutionary state. They may have  come from a range of initial masses and different 
evolutionary paths and  the origin of the instability may be different. 
The observational record is sparse because
these stars are rare and their importance has only been fully
recognized in recent years.  A greatly improved census of the likely 
progenitor class, including the most luminous evolved 
stars, the LBVs, and the warm and cool hypergiants is now needed for a complete  
picture of the final pre-SN stages of very massive stars.  
Few LBVs and hypergiants are known in our galaxy due to their rarity,
uncertainties in distance, and the infrequency of the LBV eruptions. Even the Magellanic Clouds do not provide a large enough sample to 
properly determine the relative numbers, duration, and properties of these  unstable and 
eruptive variables.
For these reasons we have begun a census of the evolved and unstable luminous star 
populations in several nearby resolved galaxies. 

In this first set of papers, we present the results of a spectroscopic survey of 
luminous and variable stars in the nearby spirals M31 and M33. In Paper I \citep{RMH13}
we discussed a small group of intermediate temperature supergiants, the warm hypergiants, 
and suggested that they were likely post-red supergiants. In this second paper, we review 
the spectral characteristics,  spectral energy distributions (SEDS), circumstellar
ejecta, and  mass loss of the LBVs, candidate LBVs, emission line
stars, and other luminous and variable stars in M31 and M33.  

Our target selection and observations with the MMT/Hectospec  and LBT/MODS1
are described in Paper I.  
We assign our targets to six different groups according to their 
spectroscopic and photometric characteristics described in \S {2}. In \S {3} we  present the evidence 
for  circumstellar nebulae, dusty ejecta, and  mass 
loss in many of these stars,  and in the last section, we summarize our results.

\section{Classification of the Stars}

Since our targets include a variety of types, for discussion we have grouped them
 by their broad spectral characteristics and other known criteria such as previous 
 variability as in the case of the LBVs.  In this section we describe the characteristics of six
  groups. The LBVs/S Dor variables demonstrate a unique spectroscopic and 
 photometric variability and are therefore discussed together as a separate group even though they share
 spectral characteristics with other groups. 
 We  discuss several  individual stars and in many cases propose an alternative interpretation
 of their characteristics. See Paper I for a detailed discussion of the warm hypergiants.
 All of the stars for which we have spectra are listed in  Table 1 in order of
Right Ascension with their  position, group type, spectral type where appropriate, and alternate 
names or designations. The M33C designation comes from an unpublished H$\alpha$ survey by Kerstin Weis.
In this paper,  we use the galaxy name and the RA of the star as its designator for brevity and to save space in the tables. For completeness, the warm hypergiants from Paper I are included
in Table 1, but are discussed only briefly in this paper. The
visual photometry from \citet{Massey06a} was cross-identified with the  near- and mid-infrared photomery 
from  2MASS \citep{Cutri},  the {\it Spitzer} surveys of M31 \citep{Mould} and M33 \citep{McQ,Thom}, and WISE \citep{Wright}. The first 10 entries are shown in Table 2 and the full table is available in 
the on-line edition. Table 2 also includes information on variability, when available, from \citet{BurgPhD} for
the M33 stars and from  \citet{Kaluzny} for M31 from  the DIRECT survey. The Burggraf compilation includes data from  several sources from $\sim$ 1920 to the present, but  most are since 1970. 
Eighteen  of our stars in M33 are in common with \citet{Clark12} and were observed at
about the same time.  The blue and red spectra in FITS format of all of the stars observed with the MMT/Hectospec are available at http:etacar.umn.edu/LuminousStars.

\subsection{Luminous Blue Variables}

Hubble and Sandage's classic 1953 paper on the brightest variables in M31 and M33
identified one object in M31 and four in M33 although Var A is now considered a post 
red supergiant, warm hypergiant \citep{RMH87,RMH06,RMH13}. 
There are now  four confirmed LBVs in M31: AF And \citep{HS,Luyten,Sharov,RMH75}, 
AE And \citep{ST,Luyten,Sharov,RMH75}, 
Var 15 \citep{Hubble,RMH78}, and Var A-1 \citep{Ros,Sharov,RMH78}. The four in M33 are 
 variables B, C, and 2 \citep{HS,Sharov,RMH75} and Var 83 \citep{vdB,RMH78}. 
Here we discuss our recent spectra of the known LBVs and related objects.

{\it AE And} is one of the more interesting LBVs. It was the visually brightest star in M33
when it was discovered in ``eruption'' \citep{Luyten} and remained at visual maximum 
for twenty years. Its emission line spectrum  closely resembles that of 
$\eta$ Car. 
It has the anomalously strong 2507{\AA} Fe II line (Szeifert at al 1996) 
which has been attributed in $\eta$ Car to a possible UV laser \citep{SEJ}.  
 In its current spectrum, several weak absorption lines are visible at  
 wavelengths below 4100{\AA}; He I $\lambda\lambda$ 4026, 4009, 3819{\AA}, N II 3995{\AA},
and a weak Ca II K line (Figure 1). N II absorption at 5666{\AA} and 5670{\AA} is also present.
These lines suggest a corresponding B2 - B3 spectral type. The earlier 2003-04 spectra  show strong H and He I emission lines that have now weakened.   
Below 4100{\AA} these lines are now in absorption.  
These changes suggest that the wind has weakened. 

\begin{figure}
\figurenum{1}
\epsscale{0.5}
\plotone{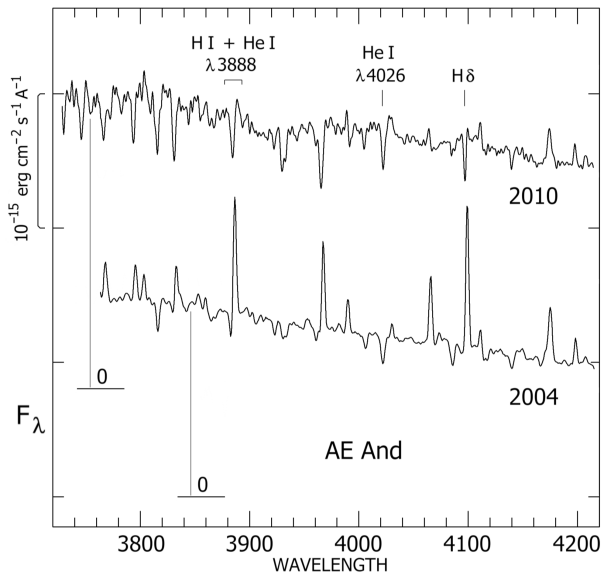}
\caption{The spectra of AE And from 2010  and 2004. Although they are shown flux
 calibrated for comparison,  the flux calibrator was observed at a
  diferent time than the 2010 spectrum. Therefore it should not be used for any absolute 
   measurements. $F{_\lambda}$  is plotted with the scale for both spectra, but
    for clarity the zero points are offset as indicated.  } 
\end{figure}

\begin{figure}
\figurenum{2}
\epsscale{0.5}
\plotone{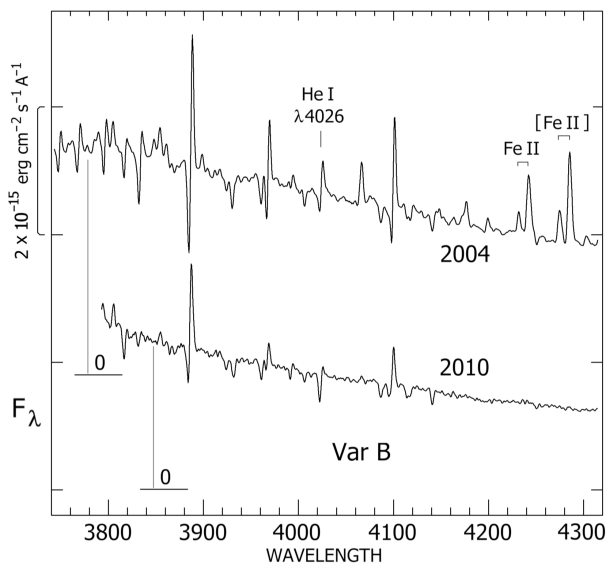}
\caption{The spectrum M33 VarB in 2010 and 2004. They are shown flux calibrated for
comparison although the flux calibrator was observed at a
diferent time than the 2010 spectrum and should not be used for any absolute
measurements.  As in Figure 1, $F{_\lambda}$  is plotted with the scale for both spe
ctra, but
 for clarity the zero points are offset as indicated. }
\end{figure}

{\it Variable B} in M33 had a high mass loss episode beginning in $\sim$ 1982, \citep{Szeif,Massey96} 
reaching an
apparent   visual brightness of $\sim$ 14.9 mag at maximum. Its eruption  
lasted until at least 1998. The observational record is incomplete after that so it is not known when the episode ended, but it had definitely begun to decline by 2002 and it had returned to its hot quiescent or minimum state in our spectrum from 2003.  \citet{Viotti} reported a visual magnitude of 17.5 for it in 2004  with a  hot emission line spectrum.  Comparison of our 2010 spectrum 
with the earlier spectrum from 2003 shows that the P Cygni absorption in the 
H and He I emission lines has weakened considerably and is now gone in the higher Balmer lines. Most of the Fe II 
and [Fe II[ emission lines are also gone in 2010 (see Figure 2). The presence of numerous absorption lines of He I and Si III and N II 3995{\AA} suggests a B2 spectral type while the O II blend  (4070-90{\AA}) and Si IV absorptions imply a slightly warmer B0 - B1 spectral type for Var B.

{\it Variable C} in M33 has exhibited a series of maximum light episodes since 
its discovery; 1940 - 1953 \citep{HS}, 1964 - 1970 \citep{Ros}, and 1982 - 1993 \citep{RMH88,Szeif}. The latter eruption was probably in decline by  1993 
as evidenced by its visual photometry and  apparent  temperature of $\sim$ 15000 K 
indicating that it was in an extended transition to its hotter state at that time \citep{Szeif}. It had 
returned to  its quiescent state by late 1998 (see \citet{Burg14}),  
but quickly entered a short  maximum 
light stage beginning in 2001 and lasting until 2005 
based on the spectra and photometry published by \citet{Viotti} and \citet{Clark12}.  
This short maximum may be similar to the brief brightening reported by 
\citet{Ros}.  Our spectrum from 2010, confirms  that Var C had returned to its hot quiescent stage. The presence of absorption lines of
N II 3995{\AA} several He I absorption lines, and S IV 4089{\AA} suggest a corresponding B1 - B2 spectral type.  Another LBV eruption began in 2011 \citep{RMH_VarC2013}. Its pre-eruption
spectrum  and a recent spectrum after the onset of the current maximum are described in
\citet{RMH2014}. 

Var C  has been highly unstable since its 1980's eruption with a shorter outburst in  2001 -- 2005, and the current brightening beginning in 2011. We suggest that Var C needs either a longer duration normal LBV eruption like the 1940's episode or perhaps a giant eruption to stabilize it in an extended   quiescent stage.

{\it Var 83} in M33 has  strong emission lines of hydrogen, He I and Fe II plus  
absorption lines of N II and  He I and the Ca II K line in absorption. 
The presence of the He I absorption lines of $\lambda$ 4026, 4009, 4144{\AA} and N II $\lambda$3995{\AA}, suggests a corresponding B2 to B3 spectral type. The Balmer lines also show asymmetric profiles 
 with very broad wings. 
 
{\it Var A-1} in M31 also shows  N II lines 
  in absorption and in addition to hydrogen and He I emission several Fe II and [Fe II] 
 lines are present. Var A-1's earlier spectrum  from 2003 showed absorption lines of He I similar to Var 83. In the current spectrum
 these are weaker and the Balmer emission lines and P Cygni profiles are much stronger.

{\it AF And} and {\it Var 15} in M31 in quiescence  have  spectra like the Of/late WN stars with strong N III emission 
lines and strong He I and hydrogen emission with P Cygni profiles. AF And's current spectrum shows little change from 2003-2004.  
Like AF And and Var 15, {\it Var 2} in M33 has nitrogen emission lines plus
He II $\lambda$4686 in emission. The spectra of AF And and Var 2 are shown in Figure 5 with other Of/late WN stars.

{\it Romano's star},  also known as GR 290 and M33-V532, \citep{Rom} is
often considered an LBV or candidate LBV \citep{Pol,Pol2011,Shol}.
It does indeed exhibit photometric variability, as much as 1.5 mag in the blue,
but it is not known to show the spectroscopic  transition from a hot star (visual
minimum) to the  optically thick cool wind resembling an A to F-type supergiant
at visual maximum that is  characteristic of the LBV/S Dor phenomenon. Instead its spectra exhibit variability
from WN8 at minimum to WN11 at visual maximum \citep{Shol,Pol2011} with a corresponding  apparent temperature range from  about 42000 K to 22000 K \citep{Shol}.
Our current spectrum from October 2010, shows He II $\lambda$ 4686 plus the N III emission lines characteristic of the late WN stars in addition to prominent emission lines of H and He I, and is consistent  with a spectrum  intermediate between its two
states. See \citet{Clark12}. Our spectrum is shown in Figure 5.

Since many LBVs in quiescence have the spectral characteristics of  late-type WN's, it is interesting to speculate on whether Romano's star is a hot star in transition to the LBV stage or if it may
be in a post-LBV state \citep{Pol2011}.

\subsection{Fe II  Emission-line Stars}

The distinguishing characteristic of this group is a blue continuum with 
strong hydrogen emission, Fe II emission  and a lack of  absorption lines which would
otherwise permit a spectroscopic classification. It must be 
remembered though that the presence of Fe II emission is ubiquitous in astronomical spectra. Fe II has  low excitation energies and is observed in many different types of stars with a wide range of luminosities and temperatures. Furthermore, the spectra of 
the stars in this group, are not all alike. Some of them have strong He I 
emission, and some have [Fe II] emission while others do not.  Those with He I emission are 
presumably hot supergiants with surface temperatures at or above 20000 K.  
[Fe II] and [O I] emission is a common characteristic of the B[e] and sgB[e] stars.   
Since not all of these stars have these emission lines, but their spectra all show Fe II emission, in this paper, we 
call them  Fe II emission line stars. \citet{Clark12} called them ``iron stars''. The notes in Table 1 indicate if 
[Fe II], He I or other lines of interest are present. 

Many of these stars also have 
significant infrared excess emission over and above what we would expect from
free-free emission from their stellar winds. Five of the  six M31 stars in this group  and six 
of the nine in  M33 have an infrared excess due to  circumstellar dust.  In a few cases an 
IR excess longwards of 8$\mu$m 
is present probably due to contaminating emission  from  surrounding H II nebulosity. The spectral energy distributions for these stars 
are discussed in the next section.   
Some of these same stars also have  emission lines of [N II] 6548{\AA} and 
6583{\AA} and [S II] 6717{\AA} and 6731{\AA}. Given the lack of or weak [O III] and [O II]
emission in their spectra, the signature emission lines of  an H II region,  
 the red [N II] and [S II] emission lines may be formed in their circumstellar ejecta and 
 can be used to estimate the  gas density. 
This special subclass of our stars are discussed in a 
separate paper by Weis et al (2014).  

Four  of these stars  are of particular interest because, like the 
warm hypergiants described in Paper I, they have 
the [Ca II] doublet in emission.
Their spectra are dominated by strong hydrogen emission with broad wings at H$\alpha$ and H$\beta$ and  some have P Cygni profiles. They all show relatively 
weak Fe II emission and no or very weak He I emission lines.
 The MODS1 red spectra of {\it M33C-15731} and  {\it M31-004417.10} confirm  that the Ca II triplet is likewise in emission and they also have the O I triplets at $\lambda$7774{\AA} and O I 8446{\AA} in emission. The latter is much stronger than the $\lambda$7774{\AA} feature likely due to $\lambda$8446 fluorescence with Lyman $\beta$.  The only apparent absorption lines in these two stars are He I 4026{\AA} and 4009{\AA} and a weak 
K-line. In addition to the [Ca II] emission, the spectra of {\it M33-013426.11} and 
{\it M31-004415.00} show weak He I emission and no obvious absorption lines. M31-004415.00 also 
has a double or split H$\alpha$ emission profile like some of the warm hypergiants. Its O I 7774{\AA} line is also 
in emission. All four stars also have an 
 infrared excess due to dust. The blue and red spectra of these four stars are shown in Figure 3.    

\begin{figure}
\figurenum{3}
\epsscale{0.4}
\plotone{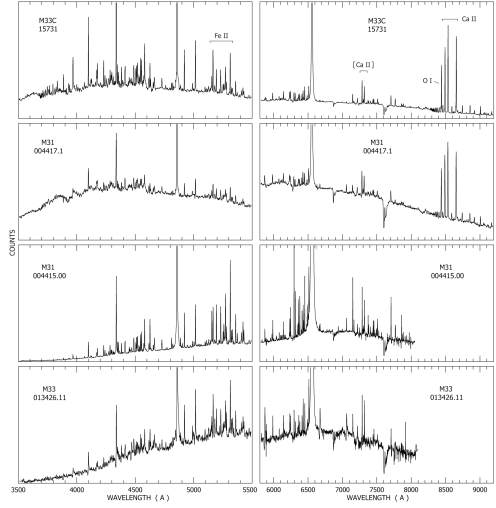}  
\caption{The spectra of the four Fe II emision line stars with [Ca II]. The two upper 
spectra were observed with the LBT and show the Ca II triplet.}
\end{figure}

\begin{figure}
\figurenum{4}
\epsscale{0.4}
\plotone{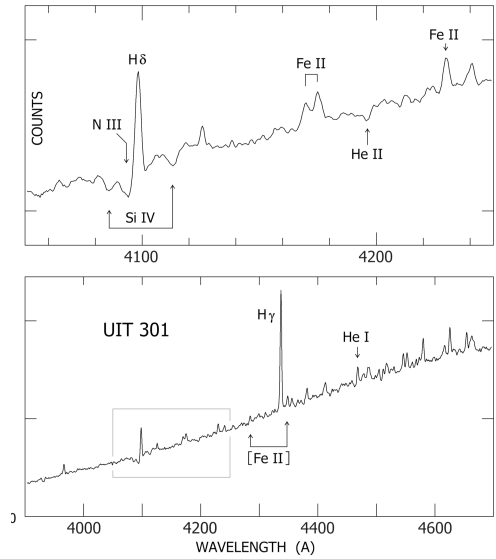}
\caption{The blue spectrum of M33-013406.63 (B416,UIT301). The upper panel show the region
around H$\delta$ with the SiIV, N II, and the He IIabsorption lines identified.}
\end{figure}

One of the Fe II emission line stars, {\it M31-004229.87}  was described as peculiar by 
\citet{Massey07}  because of an absorption line spectrum 
characteristic of an F-type star with strong Ca II H and K lines together with Fe II and 
hydrogen emission lines. We were suspicious that it might be a warm hypergiant(Paper I) and 
consequently obtained a long slit spectrum  of it with LBT/MODS1 which instead shows that  
M31-004229.87 is indeed an Fe II emission line star.
 He I $\lambda$4026 is present in absorption, and there are no F-type absorption lines and also no Ca II or [Ca II] emission. We noticed that there is a small cluster only 
about 1$\arcsec$  to the south of the star which
is the likely source of the strong H and K lines in the Hectospec spectrum in their Fig. 13 \citep{Massey07}. 

\citet{Clark12} included {\it M33-013406.63}, also known as B416 \citep{HS80} and 
UIT 301 \citep{Massey96},  
in his group of  ``iron stars''. It does indeed have weak Fe II and [Fe II] emission, 
but its spectrum also shows Si IV absorption at $\lambda$4089 and
 $\lambda$4116, N III at $\lambda$4097, and weak He II $\lambda$4200 absorption  
 plus He I emission  
 lines (Figure 4). Based on its absorption lines we  classify as a hot supergiant (\S {2.4}) with spectral type  O9.5 Ia. As we show in \S {3.1} and discuss in \S {4}, M33-013406.63 is a very 
 luminous and potentially very interesting star, although we do not consider it an LBV or LBV candidate.

\citet{Massey07} called several of these stars ``hot LBV candidates'', that is 
stars  in the hot quiescent or minimum light stage of  LBV/S Dor variables.  However, LBVs 
in quiescence  typically show  several  absorption lines  
  found  in  the spectra of normal hot supergiants, such as AE And, Var B, and Var C, or spectral 
characteristics in common with the Of/WN stars or late WN's, like Var 2 and AF And.  
Except for the hot supergiant M33-013406.63, the Fe II emission 
line stars do not share this characteristic. Given the presence of an infrared 
excess plus spectroscopic
evidence of circumstellar ejecta in many of these stars, they may indeed be candidate LBVs, 
in a post-LBV or  post-giant eruption state, or even post-RSGs since some of these
stars share some  characteristics with the warm hypergiants. Their spectral energy distributions (SEDs),
mass loss, and luminosities are presented in \S {3} and their possible relation to the LBV phase
and post-red supergiants is discussed in the last section.

\subsection{Of/late-WN Stars}

These stars have the well-known spectral characteristics of the Of and late-WN type stars with emission lines of N III and He II $\lambda$4686 in addition to strong 
hydrogen and He I emission.  The blue spectra of two late-WN stars, M33C-15235
and M31-004242.33 are shown in Figure 5 together with Romano's star (V532), and the LBVs AF And and Var 15
which share these spectral characteristics in quiescence.

\begin{figure} 
\figurenum{5}
\epsscale{0.4}
\plotone{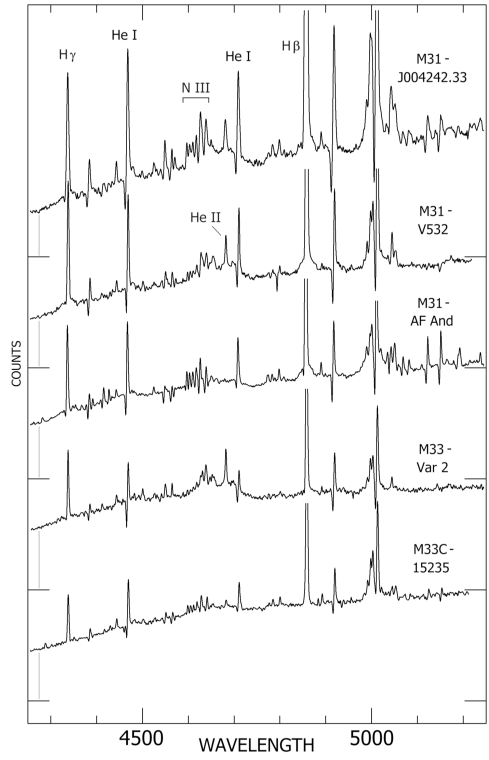}
\caption{Spectra of five Of/late-WN stars: the LBV/S Dor variables AF And 
and M33 Var2, V532 (Romano's star) in 2010, and two late-WN stars, M33C-15235 and M31-
004242.33.  }
\end{figure}

\subsection{Hot Supergiants with Absorption and Emission Lines}

This group includes  the luminous O and B-type stars, some with hydrogen and other emission lines,  with 
an absorption line spectrum that allows the assignment of an approximate spectral type.  
 Here we describe a few hot supergiants with interesting spectral characteristics, such as Fe II emission  and possible spectral variability.  

The late-O type spectrum  of M33-013406.63 with weak Fe II emission 
was described in the previous section. {\it M33C-7292}, also known as B526 \citep{HS80} and UIT 341 \citep{Massey96}, is another example of a star with the absorption line spectrum of a hot supergiant with 
hydrogen emission plus 
weak Fe II emission lines. Its spectrum  was  discussed previously by \citet{Mont} who classified it 
B2.5 Ia, but did not note the presence of the weak Fe II emission lines.  
 \citet{Clark12}  called it a candidate LBV and their brief description resembles our spectrum. 
 However,   \citet{Mont} noted that M33C-7292/B526 might be two stars based on their acquisition image. HST/WFPC2  images indeed show that it is two stars of about equal brightness 
 separated by less than 1\arcsec.       
 Massey's (2006) images  are slightly longated; the more northern component is much brighter in H$\alpha$ and is therefore the probable  source of the emission lines, but the other component is brighter in the U band. The  published spectra  are all likely composite. We include 
 it in Table 1 but do not discuss it further until spectra of the separate 
 components can be obtained. 
 
{\it M31-004425.18} has the absorption line spectrum of an early B-type supergiant, but the spectrum appears to have changed from  the one shown by \citet{Massey07} who called it a cool LBV 
candidate.  Our spectrum  shows absorption lines of He I at $\lambda$4026,4009,4144,4387 and 4471, but He I $\lambda$5015,5876, and  6678 are in emission. In addition, 
 O II$\lambda$4346-4351, Si III$\lambda$4553, and OII/CIII$\lambda$4647-51 are present in absorption as are the N II lines from  $\lambda$5666-5730{\AA}. 
On the basis of these absorption lines we classify the star B2-B3 I. The hydrogen lines are in emission with strong P Cygni absorption and  H$\beta$  and H$\alpha$ have very broad wings. The Mg II $\lambda$4481 line  in Massey's et al.'s 
spectrum (their Fig 19) is much weaker  in our spectrum. These  spectroscopic changes suggest a shift to warmer temperatures from  2006, when Massey's et al.'s spectrum  was obtained, to 2010. M31-004425.18 may be an LBV candidate, although there are no obvious Fe II emission lines in either spectrum. In the DIRECT survey the star exhibits limited variability of $\approx$ $\pm $0.1 mag.   HST images show a single star.   
This star is discussed further in \S {4.2} as a less luminous LBV/S Dor candidate. 

\citet{Clark12} have discussed the evidence for long-term  small spectroscopic variations in M33C-19725 (M33-013339.52, B517\citep{HS80}). The star also shows evidence of photometric variability on the order of $\pm$ 0.3 mag \citep{BurgPhD}. We classify it B0.5:I based on the He I,  Si IV $\lambda$4089 and 4116 plus Si III $\lambda$4553 absorption lines. However, we also note some peculiarities. For example, the He I absorption lines at $\lambda$4026, 4471 and 5876{\AA}, appear to be double due to emission in the core.
This structure is also present in H$\gamma$ and H$\delta$, although both H$\beta$ and H$\alpha$ have asymmetric profiles with P Cygni absorption plus a  second emission component on the red side. The star is embedded in a small H II region which may provide a variable contribution to the emission lines depending on the aperture size, although the nebular lines are weak.  

Our spectrum of M33C-13254 shows strong nebular emission superimposed on the absorption line spectrum  of a late B to early A-type supergiant. However this differs significantly from  its spectrum from October 2007 \citep{BurgPhD} which shows  the N III triplet Wolf-Rayet feature and He II $\lambda$4686{\AA} in emission. Earlier \citet{Conti} had classified it WNL. M33C-13254 is 
in the H II region NGC 592, a crowded field, so it is possible the 1$\farcs$5 fiber included more than 
one star, but there is no evidence of WR features in our spectrum. Its photometric 
record summarized by \citet{BurgPhD} shows an increase by one magnitude in B and V between 
2002 and 2007, and in 2011  its $B-V$ color varied from 0.2 to 0.4 mag. 
However a one magnitude increase is not sufficient to explain a transition from a WR star to a late B-type supergiant due to an LBV-like dense wind event. The expected shift in the apparent temperature would  require 
a change  in the bolometric correction of $\approx$ 2.5 mag. Furthermore there are no P Cygni profiles in the B supergiant spectrum which would be expected from  a wind. Therefore despite the close agreement between
the fiber position with the star's position in the \citet{Massey06a} catalog\footnote{Hectospec fiber  01:33:11.85,+30:38:53.69, and 01:33:11.88,+30:38:53.7(Massey)}, we suspect that this is a 
misidentification. 

These and several other stars  classified as hot supergiants are listed in Table 1 with their 
spectral types.

\subsection{Intermediate-type Supergiants (A to G-type)}  

The  intermediate-type or ``yellow'' supergiants include the visually most luminous stars in their 
respective galaxies.  Many of the most luminous A and F-type supergiants often exhibit  
hydrogen in emission  due to mass loss and winds, a characteristic  shared by the stars in this group.   A few of these stars of special interest are described here. 

Unlike  the warm hypergiants in Paper I, only one of these intermediate type supergiants, 
{\it M33-013442.14}, has an infrared excess due to dust \S {3}. Our spectrum  also  reveals what may be  a significant change from the
one published by \citet{Massey07} who called it a questionable hot LBV candidate. Instead it has the
absorption line spectrum of a late A-type supergiant with H$\alpha$ and H$\beta$ in
emission with P Cygni profiles. The O I $\lambda$7774{\AA} triplet,
a well-known
luminosity indicator in evolved stars from spectral types A to early G-type supergiants  (see \citet{AF} and references therein) is present in absorption. It does not show any  Fe II or [Ca II] emission. It is unclear
if this apparent spectral change is real though because Massey et al.'s  published spectrum  (their Fig. 17) has low resolution and low S/N in the continuum and shows only H$\alpha$ and H$\beta $ in emission plus  what look like Fe II absorption lines. We conclude that there is no evidence for a real spectroscopic change. Our blue and red spectra are shown in Figure 6.

\begin{figure}
\figurenum{6}
\epsscale{0.5}
\plotone{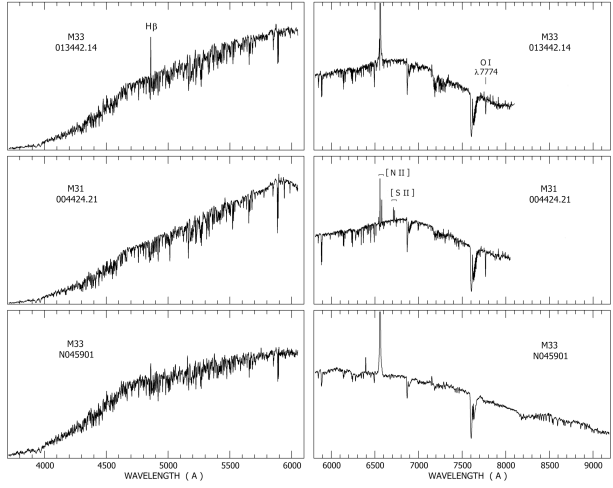}   
\caption{Spectra of the F-type supergiants discussed in \S {2.5}.}
\end{figure}

\begin{figure} 
\figurenum{7}
\epsscale{0.3}
\plotone{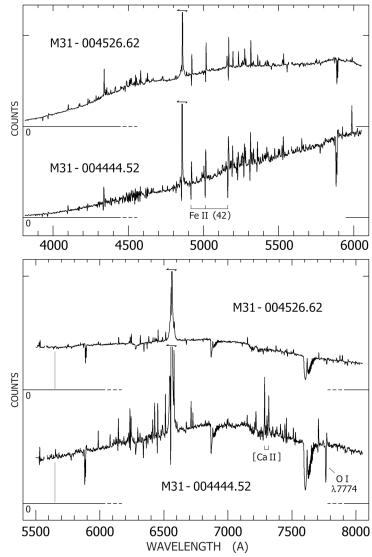}
\caption{The blue and red MMT spectra of M31-004526.62(A2eI) and the warm hypergiant
 M31-004444.52(F0 Ia).}
\end{figure}

The spectrum of {\it M31-004526.62}, with Fe II emission lines and prominent P Cygni 
profiles in the multiplet 42 Fe II  lines, closely resembles that of M31-004444.52, one of  the warm hypergiants described in Paper I. M31-004526.62  has strong hydrogen emission lines with broad wings and P Cygni profiles, [N II] emission, and the absorption line spectrum  of an early  A-type supergiant. \citet{Weis14} conclude that it has a high density circumstellar nebula, although  it does 
not have the [Ca II] emission nor  an infrared excess due to dust. 
Its blue and red spectra are shown with M31-004444.52 in Figure 7.

{\it M33C-4640} is another star with an early A-type absorption spectrum plus weak Fe II emission lines
at $\lambda$5018 and from 5100 to 5400{\AA}, but without the deep P Cygni features seen  in M31-004526.62. He I is also in emission.  

Because of our interest in the warm hypergiants  which show evidence 
for circumstellar ejecta and high mass loss events, we also observed several of 
the more luminous F-type supergiant candidates in M31 from \citet{Drout}. 
 We noticed that the O I $\lambda$7774{\AA} triplet, the strong luminosity indicator,   
 is not present in about half of the stars we observed, a fact 
that was also noted by \citet{Drout}. We  find that some of these stars are   
probable foreground stars. With their high velocities, they are likely Population II or 
Thick Disk stars in the Galactic Halo. Their  spectral types are included in Table 1.  

One of their F-type supergiants, {\it M31-004424.21}, with very strong O I $\lambda$7774
absorption, also has strong H$\alpha$ emission and  nebular [N II] $\lambda\lambda$ 
6548,6584 and [S II] $\lambda\lambda$ 6717,6731 emission, but  no [O III] or
[O II] emission.  The [N II] and [S II] emission may originate in a 
circumstellar envelope (see \citet{Weis14}). As mentioned above,  the [N II] and [S II] emission lines are present 
in several of our stars and are
possible indicators of circumstellar ejecta  discussed in the next section.
Its blue and red spectra are shown with M33-013442.14 in Figure 6. 

The paper by \citet{Drout2012} on the yellow and red supergiants in M33 was not available when we 
obtained our spectra.
Some of the M33 stars in our intermediate type group are in their list of candidate supergiants. 
One of their "red supergiants'', M33-013242.26, however is actually an Fe II emission line star.

We conclude this section by mentioning {\it N045901} in M33. \citet{Valeev10b}   suggested that N045901 
is an LBV candidate. However, contrary to their description based on a low S/N spectrum,
our spectra  show an  F-type absorption line spectrum  with
H$\alpha$ and weak H$\beta$ emission. The luminosity sensitive O I 7774 triplet is very 
weak. There are no Ca II, [Ca II], or Fe II  emission lines except for two [Fe II] lines at $\lambda$ 5158{\AA} and 7155{\AA}. Its reported small
light variations of the order of $\pm$ 0.2 mag are typical of these kinds of
stars often  referred to as $\alpha$ Cygni variability (see \citet{vG} and
references therein).  \citet{Drout2012} classify it as a supergiant, but 
they do not report an equivalent width for the O I line, one of their primary criteria
for the F-type supergiants. This is consistent with the very weak O I feature in our spectra. 
N045901's one peculiar feature is a split or double H$\alpha$ profile similar to those 
observed in some of the warm hypergiants. Evidence for doubling can also be seen  in the
weak H$\beta$ emission line. Too little information is available about this star to be 
conclusive, but we doubt that it is an LBV candidate. 
Its blue and red spectra are shown in Figure 6.  

\subsection{The Warm Hypergiants}

The warm hypergiants discussed in Paper I have A -- F-type supergiant absorption line 
spectra and strong hydrogen emission. Several of these stars also show strong Fe II emission lines and weak [Fe II].
Their spectra are  distinguished by the Ca II triplet and [Ca II] doublet in emission.  They all have significant 
near- and mid-infrared excess radiation due to free-free emission and thermal emission 
from dust. In Paper I we suggest that these evolved stars are candidates for post-red 
supergiant evolution. \citet{Kraus} described one of them, M31-004522.57, as a sgB[e] star. It does have 
weak [Fe II] emission, but with absorption lines characteristic of an early A-type supergiant, not the B-type  or veiled O-type spectrum  expected for the sgB[e] stars.  

Before leaving this section, we should mention that M33C-17194 is not included in  any of the  six groups discussed here. Our spectrum of it is  composite with hydrogen emission, absorption lines of He I like a mid to late B-type supergiant, but also with  
absorption lines like a cooler star, and  molecular band heads due to TiO beyond 5100{\AA}. 
 \citet{Massey07} classified as a WN star. We checked the HST images and find two stars only about 0$\farcs$4 apart in declination.  The eastern component is blue and the probable WN star. The other star is red.   

\section{Circumstellar Nebulae, Dusty Ejecta,  and Mass Loss}

Most of the stars included in our spectroscopic program for M31 and M33 were
selected specifically because they were known or candidate LBVs, emission 
line objects and known variables. So it is not surprising that many of them 
show spectroscopic evidence for the presence of circumstellar ejecta ranging
from nebular emission lines 
 to dust revealed by their long wavelength SEDs.

In Table 3 we list all of the stars that show one or more of the following  
indicators  for circumstellar material; [N II] emission lines, 
 Ca II and [Ca II] emission lines, and infrared excess radiation from circumstellar dust. 
Many of the hot supergiants  show weak nebular 
emission in their spectra due to contamination from emission nebulosity in the   
Hectospec aperture. 
Fortunately, the relative strengths of the [N II] lines \citep{Weis14}  allow us to separate 
a likely circumstellar nebula from an H II region.   
 The stars with H II emission contamination in their 
 spectra distinct from a circumstellar nebula are not included in Table 3. 
 H II regions also contribute PAH and dust emission at 
long wavelengths  \citep{Tielens,Peeters,Draine}. We observe this infrared emission in the SEDs of several of our stars 
embedded in emission  nebulosity which contributes to the flux especially in  the WISE 6$\arcsec$ aperture. 
In Figure 8 we show three  
examples; a normal
hot supergiant, an F-type supergiant, and the LBV AE And. Except for a small free-free contribution in the near-infrared, these three stars' SEDs show no evidence of warm  circumstellar dust but have associated nebular dust emission.  It is 
therefore necessary to distinguish 
this long  wavelength nebular emission from circumstellar dust produced by the star's 
own mass loss. 
However, many of these  stars are hot and can ionize their ejecta possibly creating a compact
region of ionized hydrogen near the star. We therefore inspected the H$\alpha$ images \citep{Massey06a}
to look for extended nebulosity at the star's position.
Combining all of this information, we distinguish the two cases, CS nebula or H II region and if    
 the star has long wavelength excess radiation, we include ``CS dust'' or possible ``H II PAH'' in the 
 comments column of Table 3. 

\begin{figure}
\figurenum{8}
\epsscale{0.4}
\plotone{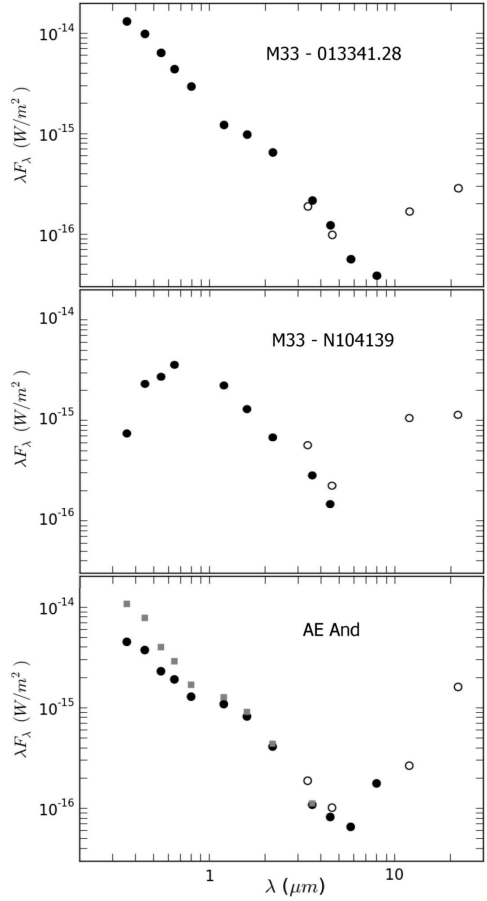}
\caption{Examples of H II region PAH and hot dust emission longwards of 8$\mu$m in the SEDs of th
ree stars; 
M33-013341.28, a normal hot supergiant (B2-3 Ia), M33-N104319, a normal F-type 
supergiant (F5 Ia), and the LBV, AE And. The observed visual, 2MASS, and IRAC magnit
udes are shown
as filled circles and the WISE data as open circles. The SEDs for the two supergiant
s are not
corrected for interstellar extinction. The extinction-corrected photomery for AE And
 is plotted as
 filled squares.} 
\end{figure}

\subsection{The Spectral Energy Distributions}

In addition to the evidence for circumstellar nebulae discussed by \citet{Weis14}, we find that  
many of the stars described 
in \S {2} have warm circumstellar dust including several of the Fe II emission line stars, 
but conclude that the known LBVs do not.  

To determine whether these stars have excess free-free emission from their stellar winds and/or 
circumstellar dust, as well as their intrinsic luminosities, we must first correct their SEDs for interstellar extinction. This is an 
uncertain procedure  for stars with strong emission lines. Their broadband colors cannot be safely used especially in the blue-visual region. We therefore estimate the visual extinction, A$_{v}$, from other indicators including the well-known  
relation between the neutral hydrogen  column density (N$_{HI}$) and the color excess, E$_{B-V}$ \citep{Knapp,Savage},
and A$_{v}$ from the Q-method for nearby OB-type stars  within a few arcsec of the target 
assuming that their UBV colors from \citet{Massey06a} are normal. We use R $=$ 3.2 and the standard 
extinction curve from \citet{Cardelli}. The results from these different methods are summarized in Table 4;    
A$_{v}$ defined as the foreground A$_{v}$ ($\approx$ 0.3 mag) plus 1/2 A$_{v}$\footnote{Since we do not know the exact location of the stars along the line of sight with respect to the neutral hydrogen, we adopt half the extinction derived in this way.} from N$_{HI}$ from the recent H I surveys 
of M31 \citep{Braun} and M33 \citep{Gratier}, A$_{v}$ for several of the known LBVs 
from \citet{Szeif} who likewise used N$_{HI}$ from earlier H I surveys,  
A$_{v}$ from the observed colors for the F-type supergiants, and A$_{v}$ from nearby stars with the number of stars used
in parenthesis. We favor the extinction estimates from nearby stars when available because of 
their proximity to the target stars, compared to  the H I surveys which have  spatial resolutions of 30$\arcsec$ and 17$\arcsec$ for M31 and M33, respectively. Table 4 includes the adopted A$_{v}$ and 
their corresponding  
extinction-corrected visual and bolometric luminosities with  the same distance moduli used in Paper I, 24.4 mag for M31 and 24.5 mag for M33.

The SEDs for the known LBVs are shown in Figure 9; AE And is in Figure 8. LBVs are variable and 
understandably  the broadband visual, and near- (2MASS) and mid-infrared (IRAC and WISE)
photometry was not all observed at the same time, and therefore do not always yield SEDs corresponding to the same time frame.  This is especially true for 
Var B and Var C which as noted in \S {2.1}, have experienced LBV-type optically thick wind episodes
in the past decade when these data were obtained. The visual photometry for Var C from \citet{Massey06a} 
is  consistent with  earlier B and V magnitudes \citep{Szeif}.  We therefore adopt the  corresponding JHK 
photometry from \citet{Szeif}  for its  SED in Figure 9.  Var B is not shown here for this reason; see
Figure 9 in \citet{Szeif} for its SED near its visual maximum. Interestingly, comparison with the earlier photometry for M31-Var 15, shows that it  declined by about one magnitude from 1992 to 2001. The visual and near-IR photometry used here for the other LBVs are consistent with the earlier 
observations. As we noted in \citet{Szeif}, 
the LBVs show excess near-IR radiation due to free-free emission. With the addition of 
the IRAC and WISE bands, we 
conclude that {\it there is no evidence for hot or warm  CS dust in the LBVs} based on their SEDs in Figure 8 and 9. The SEDs of AF And, Var 15 and AE And show an increase in flux at and longwards of 8$\mu$m  but no evidence of warm circumstellar dust at the shorter wavelengths. We attribute this apparent infrared excess 
to probable  
 PAH  and dust emission from  H II nebulosity primarily in the 6$\arcsec$ WISE aperture. AE And could be a good candidate for an associated  compact H II region, but  close inspection of the H$\alpha$ image of AE And shows a wisp of nebulosity passing through the position of the star, although it also has a dense N-rich circumstellar  nebula \citep{Weis14}.

\begin{figure}
\figurenum{9}
\epsscale{0.6}
\plotone{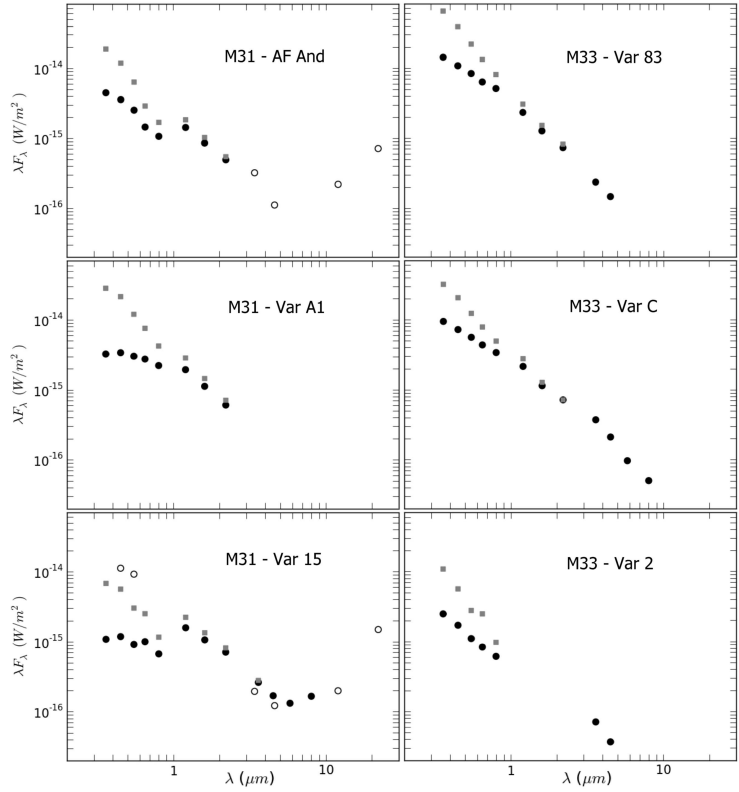}
\caption{The SEDs for the confirmed LBVs. The symbols are the same as in Figure 7, e
xcept for Var. 15. We  
also use open circles for the earlier B and V photometry when it was brighter. As no
ted in the text we've used visual and near-infrared photometry for \citet{Szeif} for
 Var C. Note that AF And and Var. 15 show PAH emission.}
\end{figure}

\citet{Szeif} combined  UV fluxes from HST/FOS near-UV spectroscopy with visual and near-IR 
photometry for  several of the M31 and M33 LBVs
to estimate their corresponding bolometric  luminosities and temperatures from 
their SEDs.  We therefore adopt the A$_{v}$'s from  that paper  and their derived 
bolometric luminosities, adjusted for the small change in distance moduli, for AE And,  AF And, 
Var B, Var C and Var 83 in Table 4.

Figures 10a and 10b show the SEDs for the Fe II emission line stars in M31 and M33 
with long wavelength evidence for CS dust.  
Our criterion for concluding that these stars  have circumstellar dust is based on a rise 
in their SEDS $>$ 2$\mu$m above what would be expected from an extrapolation of their 
visual SEDs or 
a  contribution from free-free emission at the longer wavelengths. Due to their faintness, 
JHK photometry from 2MASS is not available for some of them, but nevertheless, their SEDs 
clearly show an excess at wavelengths longwards of 3$\mu$m.  The SEDs for many of 
these stars also show an excess in the U band. We note that the observed U-B colors of many 
of the 
Fe II emission line stars are unusually negative, that is very blue, and even moreso when 
corrected for the adopted A$_{v}$'s.  Interestingly, this is a characteristic that they 
share with the LBVs and 
warm hypergiants. The broadband U magnitude is much less affected by emission lines in their 
spectra than the B magnitude wavelength range where
there are numerous strong Fe II emission lines. The apparent U-B color excess is thus 
due mostly to increased flux in the U band as seen in their SEDs.  
Given the very strong hydrogen emission lines in their spectra, we suggest that this excess U band
 flux is due to emission in the Balmer continuum below 3600{\AA}.
The R band flux is also elevated in several 
of the SEDs due to the very strong H$\alpha$ emission line. The SEDs for M33C-15731(UIT 212)
and M33-013406.63(UIT 301) which have additional NUV and FUV flux measurements \citep{Massey96} are shown in Figure 10c.  

\begin{figure}
\figurenum{10a}
\epsscale{0.4}
\plotone{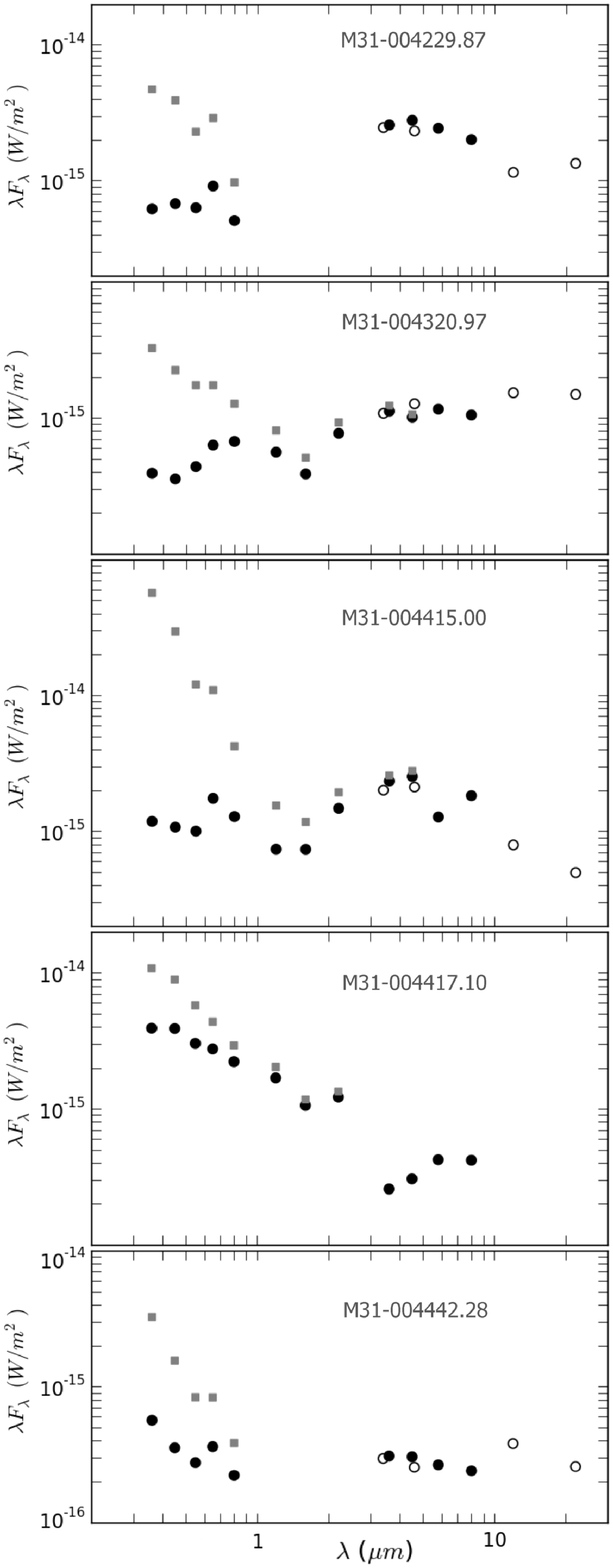}
\caption{The SEDs for five Fe II emision line stars in M33 with circumstellar dust. 
The symbols are
the same as in Figure 7.}
\end{figure}

\begin{figure}
\figurenum{10b}
\epsscale{0.4}
\plotone{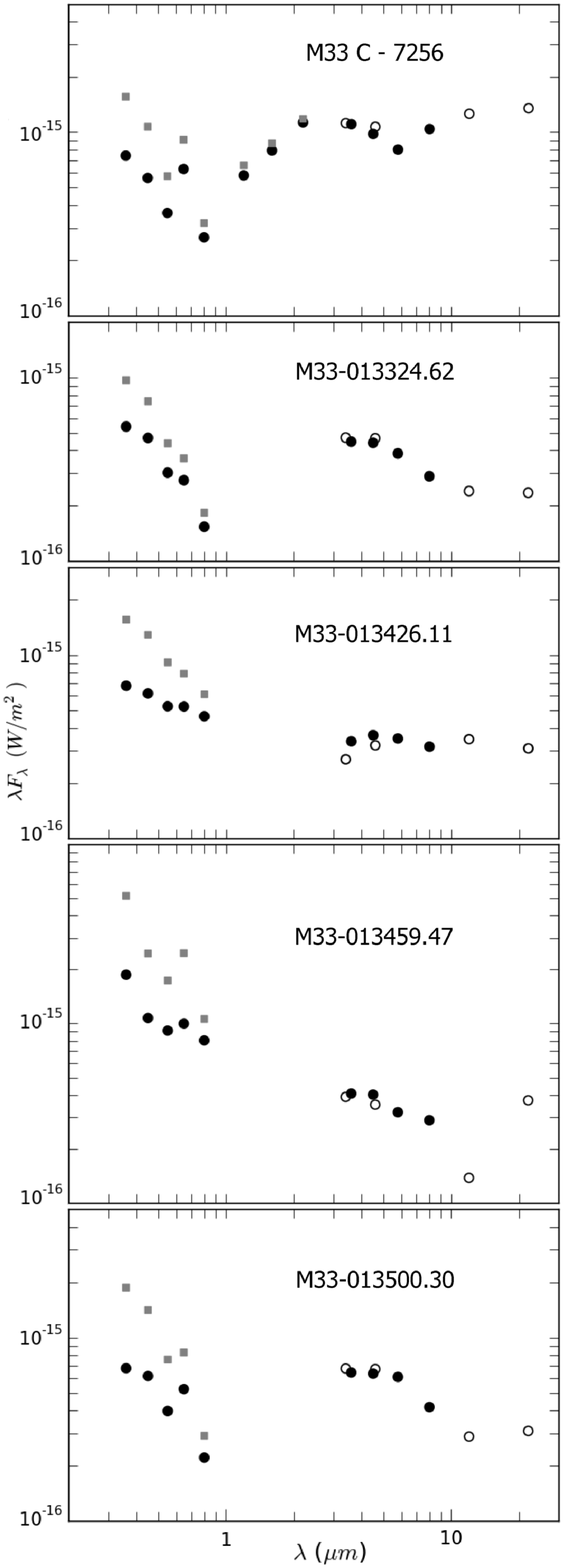}
\caption{The SEDs for five Fe II emision line stars in M33 with circumstellar dust. 
The symbols are 
the same as in Figure 7. Although M33-013459.47 shows the characteristic PAH upturn,
 it is shown
 here because its SED shows an excess at 3 to 8$\mu$m; its infrared excess is very li
 kely due to a combination of free-free
 emission, some CS dust with a contribution from HII region PAH emission.  }
\end{figure}

The long wavelength SEDs of the 
Fe II emission line stars are complex including free-free emission and  scattered light from warm dust in the near-infrared, and thermal emission longwards of 3.5$\mu$m.
M33C-15731 in Figure 10c is a good example. We used the LMC average  extinction curve 
from  \citet{Gordon} to correct its NUV and FUV fluxes. Earlier work (\citet{Szeif} and references therein) had found that the LMC UV extinction curve was more appropriate for the massive stars in M31 and M33.  The addition of the short wavelength fluxes allow us to estimate a temperature based on a  21,400K  blackbody  fit through its UV and visual
fluxes. Its SED also   illustrates the U band excess, the near-IR free-free emission, and additional 
excess radiation at the longer wavelengths. 

M33-013406.63(B526/UIT 301) is embedded in a prominent  H II region. Its SED, also shown in Figure 10c, has a significant excess at long wavelengths most likely due to PAH emission  from  the surrounding H II region,  but  no evidence for hot or warm  circumstellar dust. Based on its O9.5 Ia spectral type we would expect a temperature near 30,000 K \citep{Martins,Massey04}. With the adopted A$_{v}$ of 0.7, however it is clear that a 30,000 K 
blackbody is not a good fit to the SED, and a significantly lower temperature 
$\approx$ 20,000 K is implied. The discrepancy is most likely due to  
 the uncertainty in the visual extinction correction. 
 The SED is more consistent with the expected temperature with a higher A$_{v}$ of 1.2 mag, the maximum value from  the   neutral hydrogen column density.   
An A$_{v}$ as high as  1.5 mag is even suggested based on its observed B-V color, although the color may be
affected by its strong hydrogen emission.
With this possible range in the visual extinction, M33-013406.63's visual luminosity is
M$_{v}$ is -9.1 to -9.6 mag and with a bolometric correction of -3.1 corresponding  to is apparent spectral type, its bolometric luminosity is -12.2 to -12.7 mag, exceeding the 
luminosity of $\eta$ Car.  M33-013406.63  is very likely a binary or multiple star, but in any case, it is a very luminous star and is  discussed further in \S {4}. 

\begin{figure}
\figurenum{10c} 
\epsscale{0.50}
\plotone{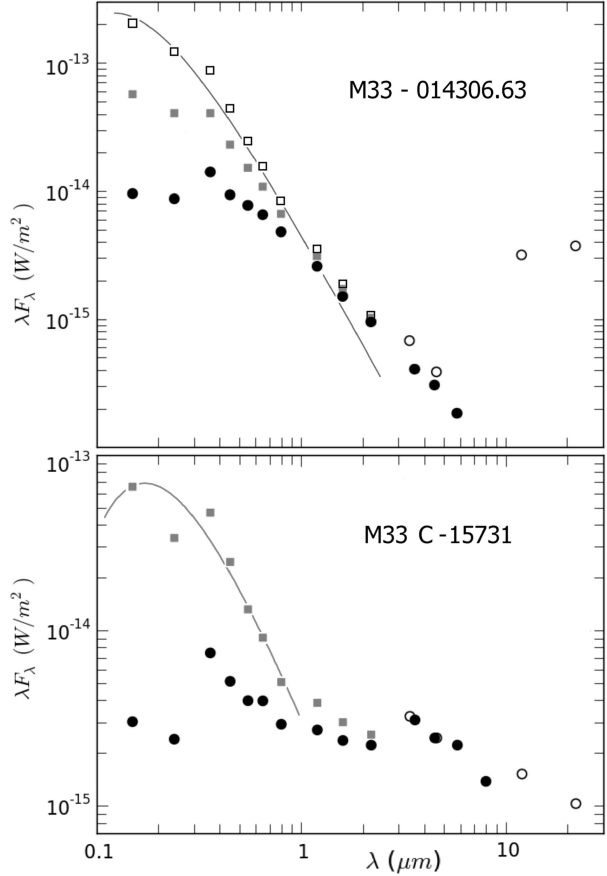}
\caption{The SEDs for M33C-15731 and M33-013406.63. The symbols are the same as in F
igure 7 with the addition of open squares for the  second extinction correction for 
M33-013406.63 discussed in the text. 
The NUV and FUV fluxes are 
from the 1500{\AA} and 2400{\AA} magnitudes from the UIT survey \citep{Massey96} and
 transformed to fluxes following the prescription in that paper. They were corrected
  for interstellar extinction using the
  adopted A$_{v}$ and the average extinction curve for the LMC for these UV wavelength
  s \citep{Gordon}.
  Like the SEDs for the LBVs \citep{Szeif}, they show the dip at 2000 -- 2400{\AA}, an
  d have an
  excess in the U band most likely due to Balmer continuum emission. A 21,400K blackbo
  dy is shown fit 
  through M33C-15731's blue-visual and 1500{\AA} broadband points. The near-IR free-fr
  ee emission and the 
  presence of dust $>$ 3.5$\mu$m are apparent. M33-013406.63 is shown corrected for va
  lues of A$_{v}$, 0.7 mag and 1.2 mag discussed in the text. M33-013406.63 shows free
  -free emission in the near-IR, the PAH signature, but no circumstellar dust.  A 30,0
  00K  blackbody is shown through the upper points.} 
\end{figure}

None of the other Fe II emission line stars in M33 have UV fluxes from the UIT survey. We identified four of 
the M31 stars in the GALEX survey,  but only one, M31-004442.28, was measured in both the 
NUV and FUV bands. 

Without temperature indicators in their spectra together with the uncertainty 
in their extinction corrections, the bolometric corrections are uncertain  for the other 
 Fe II emission line stars.  Except where noted, their bolometric luminosities                                     listed in Table 4 are based on integrating 
their SEDs from 0.35 to 22$\mu$m.. Five  of these stars  
show a significant contribution to their SEDs from circumstellar dust and to their 
total luminosities, even when we consider the possibility that the visual 
extinction corrections may have been underestimated and the stars are actually visually 
brighter. For the other stars, the dusty contribution is at a relatively 
low level.  In all cases the circumstellar dust appears to be fairly warm. Longwards of 
2 -- 3$\mu$m, the relatively flat SEDs suggest a range of dust temperatures  
from a few 100 to 1000K.  Based on their visual SEDs and the presence of 
He I emission
in some of their spectra, many of these stars are fairly warm, but without information on their 
UV fluxes the bolometric luminositites in Table 4 are  very likely under 
estimates for most of the Fe II emission line stars. For M33C-15731 we assumed 
the  temperature from  the blackbody fit (Figure 10c) plus the contribution from  free-free 
emission and dust, and for M33-013406.63 we use the inferred temperature and bolometric correction corresponding to its spectral type with the range of extinction corrections discussed above.

The SEDs for three intermediate-type supergiants discussed in \S {2.5} are shown in Figure 11.
The adopted visual extinction corrections for these stars is based on their observed colors. 
Their blue-visual colors are not seriously compromised by the emission lines in their spectra, 
and the A$_{v}$'s from the different methods in Table 4 are consistent, although  for M33-013442.14 only one nearby star is available for comparison, and the maximum A$_{v}$ from 
N$_{HI}$ would be  2.1 mag, equal to the A$_{v}$ from its colors. 
It also has CS dust which could contribute to its extinction. 
We used  the bolometric correction appropriate to the  
spectral types for the two stars in M31  and integrated the SED for M33-013442.14 even though 
the contribution from  its long wavelength excess is small.  We note that all three 
of these stars which show evidence for CS nebulosity, ejecta, and winds are quite luminous. 

We mention N045901 here and include it in Tables 3 and 4 because it has been proposed as a possible LBV
candidate but as we emphasized in \S {2.5} that suggestion is not supported by its spectrum. 
Cross-identification with possible 2MASS and IRAC sources is also doubtful, see the notes to 
Table 2. Thus there is no convincing evidence for circumstellar ejecta.   

\begin{figure}
\figurenum{11}
\epsscale{0.5}
\plotone{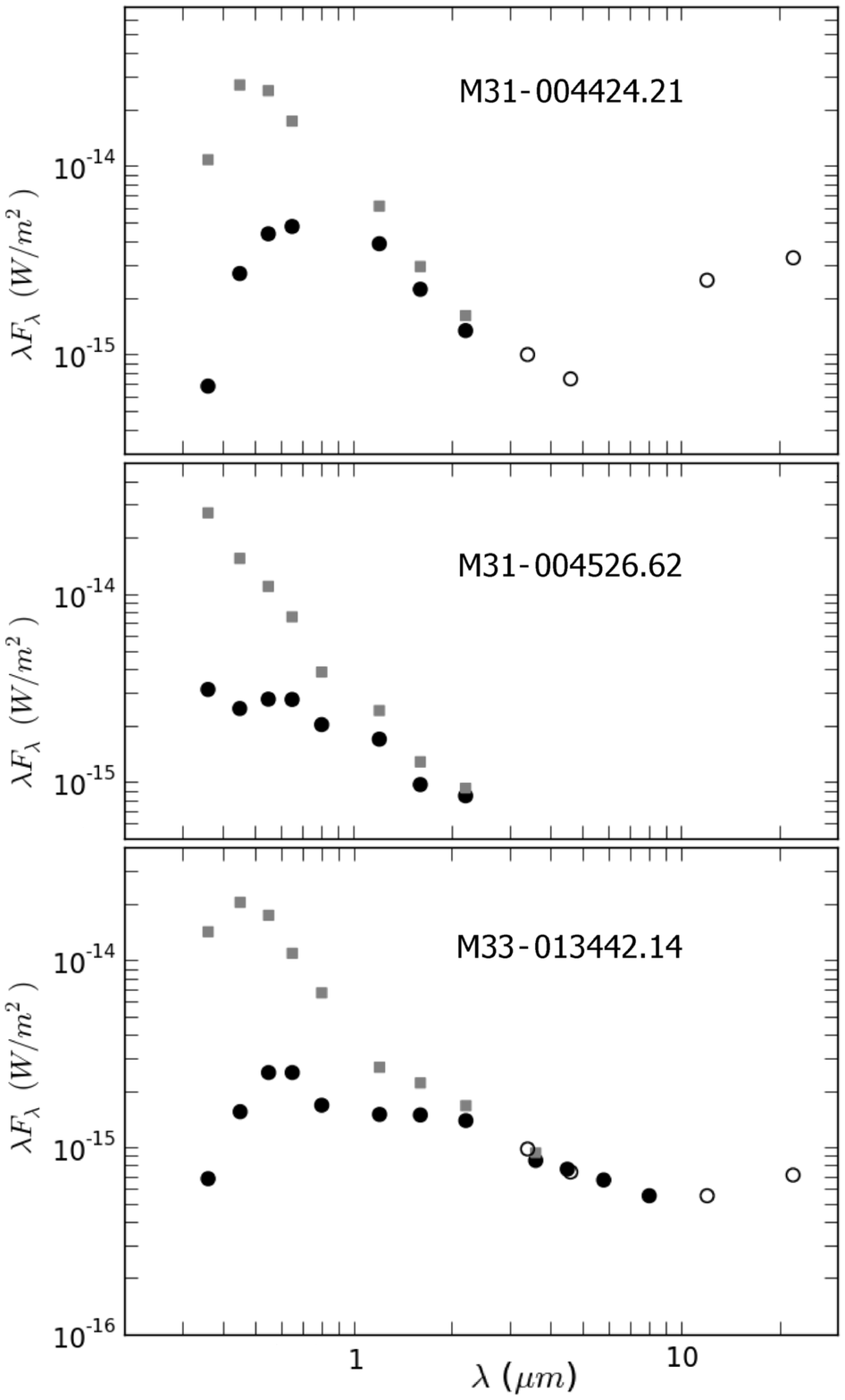}
\caption{The SEDs for three of the luminous F-type supergiants. M33-013442.14 has exces
s radiation from both free-free emission in the near-infrared and from dust at the long
er wavelengths. M31-004424.21 shows PAH emission.}
\end{figure}

\subsection{Circumstellar Nebulae, Winds, and Outflows} 

P Cyg profiles, asymmetric profiles, broad wings, and double or split hydrogen emission lines, are all 
indicators of winds, mass loss and outflows. Most of the stars in our program, the hot supergiants, 
the WR stars, as well as the LBVs, show prominent P Cygni profiles in the hydrogen emission lines. 
These stars all have winds and mass loss. 
Consequently, in this section we will focus on the evidence for circumstellar nebulae and gaseous outflows in the LBVs, 
the Fe II emission line stars, and some of the intermediate-type stars of interest.

Using the ratio of the [N II] nebular emission lines $\lambda5755/\lambda6584$, 
in the spectra of our luminous stars, \citet{Weis14} found that most  of the known LBVs, several
candidates, and Fe II emission line stars have electron densities 
on the order of 10$^{6}$ cm$^{-3}$ or higher, much greater
than for an H II region, suggesting the presence of circumstellar nebulosity. 
These stars are included in our Table 3 with CS nebula in the remarks column. Three of the four
Fe II emission line stars with [Ca II] in emission (\S {3.2}) also have  [N II] 
ratios indicative of a circumstellar nebula.
For comparison, the [Ca II]/Ca II ratio for  M33C-15731 and M31-004417,  yield higher densities of 5 -- 6 $\times$ 10$^{7}$ cm$^{-3}$ which are also somewhat
 higher than we found for the warm hypergiants in Paper I.

It is customary to measure the wind speed or terminal
velocity ($v_{\infty}$) from the blue edge of the P Cygni profile, but for these moderate 
resolution spectra, and especially for the stars with poorer S/N and poorly defined 
continua, the absorption minimum  provides a
 more well-defined differential measurement.
The stars with P Cygni profiles in their hydrogen, He I, and/or Fe II emission lines are listed in Table 5 with their outflow velocities measured from the absorption
minima in their P Cygni profiles relative to the emission line peak. The outflow velocities 
from the  double H emission lines when present in a few stars are also given.   
 For those spectra with good S/N, such as Var C in M33,  the results from the absorption minimum  are consistent with and parallel the behavior of the terminal or blue-edge velocity \citep{RMH2014}. 
For comparison, the wind speeds from  the P Cygni profiles in the normal supergiants in M31 and M33 measured the same way are also included in Table 5. 

LBVs are well-known to have low wind speeds of 100 - 200 km s$^{-1}$ during their 
eruptions or maximum light phase. We find that the known LBVs in M31 and M33 also have relatively 
low outflow velocities even in their quiescent phase.  
 The outflow velocities of these LBVs in quiescence are typically 50 - 100 km s$^{-1}$ less than
for the normal supergiants in our sample with similar spectral types. 
For example, the outflow velocities of the LBVs with early B-type spectra in their quiescent 
or hot state average $-$170 km s$^{-1}$    compared to $-$220 km s$^{-1}$  for the 
normal supergiants.  
 Similarly,  the three LBVs with Of/WN type spectra have wind velocities   
 typically 100 km s$^{-1}$ less than the average for the Of/WN stars 
in our sample. GR290, Romano's star, shows the same difference.  
Although wind speeds are expected to have some dependence on metallicity, we do not find any 
significant difference in the wind 
velocities between the M31 and M33 LBVs  in this small sample or betwen the normal supergiants in the two galaxies. This is not surprising given the range in the measured velocities and their associated 
measured uncertainties.

The well-studied Galactic LBV, AG Car, similarly has a relatively low terminal  velocity during its quiescent or hot stage when it  is spectroscopically like an Of/late WN-type star 
(WN11 \citep{LSmith}). Its terminal velocity, measured from the blue-edge \citep{Leith,Stahl,Groh},   is  250 to 300 km s$^{-1}$.  \citet{Groh} reported an even  lower velocity of 105 km s$^{-1}$ 
during its 2001 minimum  or quiescent state, although it had not returned  to it previous high temperature. These wind velocities are low compared to terminal velocities of 400 - 1300 km s$^{-1}$ for the Of/late-WN stars and WNL stars (Crowther,Hillier \& Smith 1995a,b) 
and 400 to 1000 km  s$^{-1}$ for  Galactic and LMC early-type supergiants  \citep{Crowther,Mokiem}.  

{\it Thus the winds of the LBVs even in their hot, quiescent state are apparently 
slower and presumably denser than the normal hot supergiants with similar 
spectroscopic temperatures.} 
This should not be surprising. LBVs are presumably close to their Eddington limit \citep{HD94}. They have lost much of their mass in previous S Dor-type maxima or even in a previous giant eruption. Thus their effective gravities and escape
velocities are now much lower.

\subsection{Circumstellar Dust and Mass Lost Estimates}

The  3.5$\mu$m to 8$\mu$m fluxes from the IRAC data clearly show the presence of dusty 
circumstellar ejecta  in the SEDs for several of stars discussed in this paper, and  
the WISE data indicate  that the thermal emission  extends to even longer 
wavelengths in several cases. We can  estimate the mass of the circumstellar ejecta from the flux at mid-infrared 
wavelengths using the equation for the dust mass and following the prescription in Paper I 
with  the flux at 8$\mu$m  and  an assumed grain temperature of 
350 K at that wavelength. The results are in Table 6.  

Overall, the mass in the ejecta ranges from a high of about two times 10$^{-2}$ to a few times 
10$^{-3}$ M$_{\odot}$. The highest  values for the Fe II emission line stars  are  comparable to our  estimates for the warm hypergiants 
in Paper I, but overall, they average somewhat lower.  
   
\section{Summary and Concluding Remarks} 

In this section we summarize our results for the LBVs, other variables, the Fe II emission line stars, the intermediate-type  supergiants with comments on post-red supergiant evolution, and the very luminous  M33-013406.63 (B416, UIT301).  The stars discussed in this section are shown on an HR Diagram in Figure 12.

\subsection{The Luminous Blue Variables}

{\it We find that the confirmed LBVs  have  low wind speeds} in their hot, quiescent or visual 
minimum state, compared to the B-type supergiants and Of/WN stars
which they spectroscopically resemble. In the case of Var C, currently in eruption \citep{RMH2014}, there is little difference in the outflow velocities between the two states.
The lower wind speeds and presumably denser winds, even in quiescence, may be another distinguishing property  of the LBV/S Dor variables and consequently may  aid in identifying 
candidates in addition to the characteristic, but infrequent,  S Dor variability. 
This may be a significant clue to the structural trait that causes the LBV instability. Due to their record of higher mass loss episodes, LBVs may be close to their effective Eddington limit which will result in  lower escape velocities and outflow speeds.  

Although many of  the confimed LBVs have  circumstellar nebulosity \citep{Weis14}, we find no evidence 
for associated hot or warm dust although we can't rule out the possibility of cold dust at much longer wavelengths. 
\citet{Oksala}  emphasized that the LBVs in the Magellanic Clouds also lack hot dust
and \citet{Kraus} concluded the same for two  LBVs in M31. Most of the  confirmed LBVs in the LMC and 
SMC also show a lack of of warm dust in the 3.5 to 8$\mu$m region (see Bonanos et al. 2009, 2010).   

In addition to their unique spectroscopic and photometric variability, LBV/S Dor variables in quiescence or their hot, minimum light stage,   define or lie on the ``LBV/S Dor instablity strip'' in the upper 
HR Diagram \citep{Wolf89,HD94}. 
Most stars that lie in this locus on the HR Diagram  however are {\it not} LBVs. Normal hot supergiants 
are also found in the S Dor  strip. 

The Of/WN star, UIT 008 in M33 has an unusually slow wind speed (Table 5) for this type of star, and thus may be an example of star that has shed a a lot of mass. Its very high temperature corresponding to its O7 -- O8 spectral type of $\approx$ 35,000 K \citep{Martins,Massey04} and bolometric luminosity of -10.7 place it near the top of the S Dor instability strip near stars like AG Car and AF And. On this basis, it may be a candidate LBV and a star worth watching although it is not a known variable.

\begin{figure}
\figurenum{12}
\epsscale{1.0}
\plotone{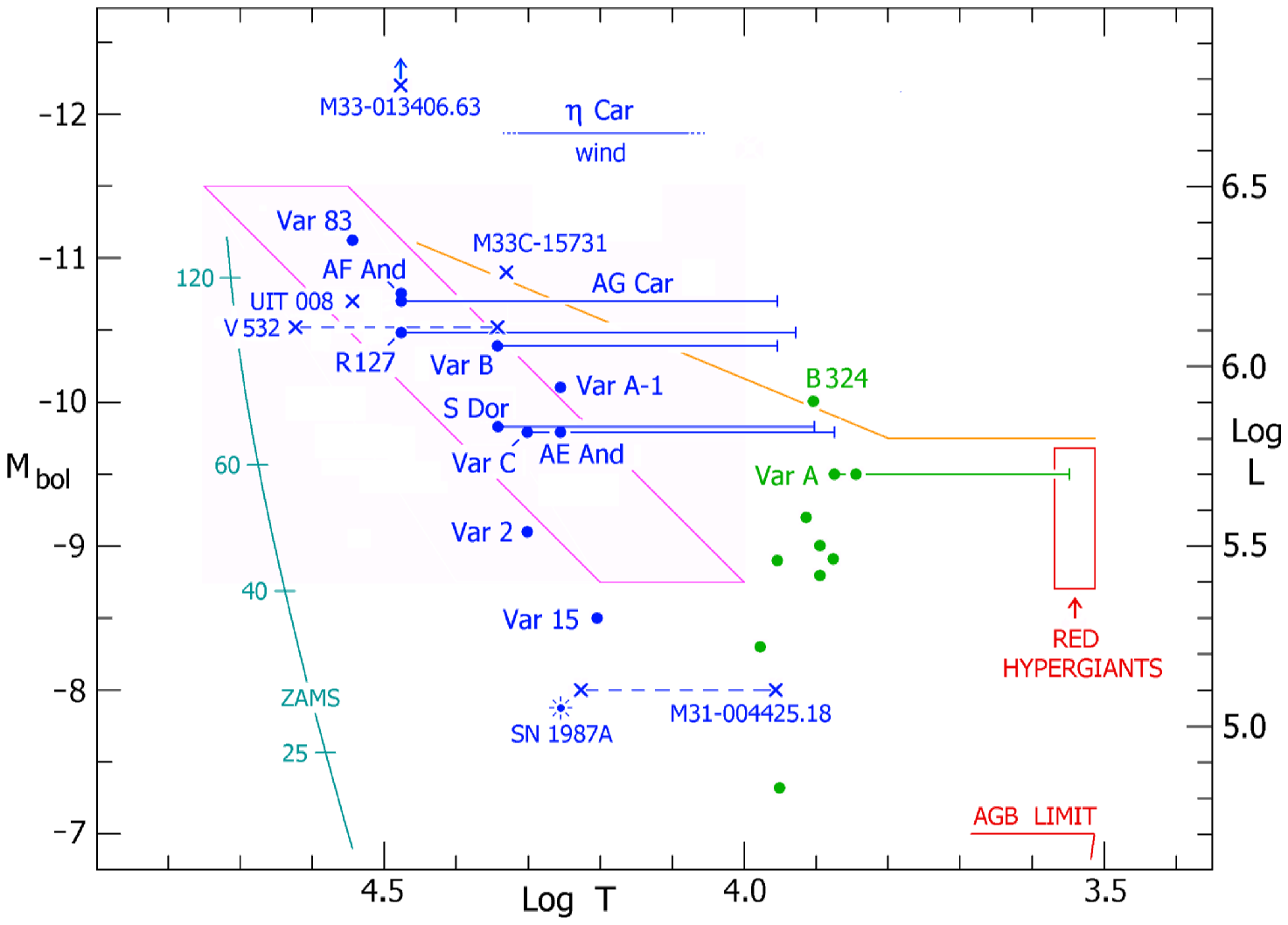}
\caption{A schematic upper HR Diagram  with the locations of the confirmed LBVs (blue d
ots), the warm hypergiants (Paper I) and candidate post-red supergiants (green dots), a
nd other stars discussed in \S {4} (blue $\times$'s).  The outline of the LBV/S Dor ins
tability strip (pink) and the empirical upper luminosity boundary (gold) \citep{HD94} a
re also shown. The LBV/S Dor transits to the cool dense wind state are shown as solid b
lue lines and those for V532/GR290 (Romano's star) and M31-004425.18 are dashed blue li
nes. The positions of $\eta$ Car and the well-studied Galactic and LMC LBVs, AG Car, S 
Dor, and R 127 are shown for comparison. The apparent temperatures of Var 2, Var 15 and
 Var A-1 are estimates and their positioms on the HRD  are uncertain. Note that the sol
 id blue transit line for Var C passes through the position of AE And.} 
\end{figure}

\subsection{Other Variables}

Romano's star (GR290/V532) is often called an LBV, but it does not exhibit S Dor transits to the cool, dense wind state.
Instead, it  varies between two hot states on the HR Diagram (Figure 12) \footnote{See Table 4 for its 
extinction and luminosity estimate.}  characterized by WN spectroscopic featues. 
It also shares the lower wind speeds of the LBVs (Table 5). Given its shared characteristics with the LBVs plus its strong WR spectroscopic characteristics, we propose that Romano's star is likely in a post- LBV/S Dor state also suggested by \citet{Pol2011}.    

In \S {2.4} we suggested that M31-004425.18 may be an LBV candidate based on the evidence for a 
spectroscopic
transition from an apparent early A-type supergiant to  an early B-type. Additional confirming observations will be needed, although like UIT 008, M31-004425.18 also has a slow wind of -153 km s$^{-1}$ for an early B-type supergiant (Table 5). 
We've also estimated its luminosity (Table 4)  with an extinction correction 
from the neutral hydrogen column density.  Adopting the photometry from \citet{Massey06a} and assuming 
that it applies to the early A-type spectrum, we find an M$_{v}$ of -7.7 mag and M$_{Bol}$ of -8. Its position is shown on the HR Diagram  (Figure 12) with its  apparent shift to the warmer early B-type spectrum. M31-004425.18 has a rather low luminosity compared to the other LBVs and even falls below the lower limit of the less luminous LBVs on the HR Diagram.

\subsection{The Fe II Emission Line Stars}

One of the questions we wanted to address is the possible relation of the luminous Fe II emission line 
stars to LBVs, or
to the LBV stage, and to the warm hypergiants.  Based on the 
available data, the confirmed LBVs in M31 and M33 do not have significant circumstellar dust. 
 In contrast, several of the Fe II emssion line stars have both circumstellar 
dust and nebulosity indicative of past or present  mass loss (Tables 3 and  6). Unlike LBVs in quiescence their spectra do 
not  have absorption lines typical of hot supergiants or late WN-type stars.   

M33C-15731(UIT 212) is the most luminous Fe II emission line star. Its luminosity of M$_{bol}$ $\approx$ -10.7 mag (Table 4) places it the realm of the evolved hot supergiants on the upper HR Diagram.  It also has a circumstellar nebula \citep{Weis14}.
With our temperature estimate from  its SED (Figure 10c) of $\approx$ 21,000 K, M33C-15731 will fall  to the right of the S Dor instability strip. Given the uncertainties  in these
estimates, it thus may qualify as a candidate LBV except for its dusty ejecta not shared
with the known LBVs.

As discussed in \S {3.1}, without absorption lines in their spectra or other information, the temperatures of most of the Fe II emission line stars are unknown. Our estimates of their luminosities (Table 4), however shows that
most of them  lie below the constant luminosity portion of the upper luminosity boundary for evolved stars in the HR Diagram (Figure 12). They have luminosities similar to the less luminous LBVs \citep{HD94}. Several have  dusty circumstellar ejecta. The Ca II and [Ca II] emission lines in four of them, is a characteristic shared with the warm hypergiants (Paper I) and with some B[e] supergiants, although not all of our Fe II emission line stars have [Fe II].     
 \citet{Aret}  reported  Ca II and [Ca II] emission in the spectra of eight B[e] supergiants in the Magellanic Clouds, and noted their spectroscopic similarities to the 
yellow hypergiants. They  suggested that those in the less-luminous regime may be candidates
for blueward evolution from the yellow hypergiant phase.  

\citet{Oksala} concluded  that the C$^{12}$/C$^{13}$ ratio in the CO emission 
features, in many 
 sgB[e] stars support an evolved,  post-main sequence state in a likely pre-red supergiant stage. However,   the less luminous ones and several of the high  luminosity sgB[e] stars do not show the CO emission feature, leaving their evolutionary state uncertain. In a recent review  of high angular  
resolution observations of know sgB[e], \citet{deWit}  conclude that the stars in their sample have rotating 
equatorial disks, and several are identified with short-period binaries. They favor a pre-red supergiant 
phase for the sgB[e] stars.  Although the evolutionary state of the sgB[e] stars is still 
somewhat ambiguous, there is concensus that they are in post-main sequence state and  at least some of them  are apparently pre-red supergiants and therefore probably 
not related to the LBVs.  Given that Fe II emission is so common among the luminous stars, this situation
is not surprising.

\subsection{The Intermediate-Type  Supergiants and Post-Red Supergiant Evolution}

Blue to red evolution across the upper HR Diagram, from the end of the main sequence to the 
red supergiant stage  
and  possibly back to warmer temperatures is expected to be relatively rapid. Consequently, 
there are not many known luminous yellow or intermediate-type supergiants. Furthermore, there 
is increasing
evidence that massive stars above about 20M$_{\odot}$ may not end their lives as Type II
supernovae in the red supergiant stage, but instead evolve back across the HR Diagram to
warmer temperatures before the terminal explosion, see Paper I. 
Criteria for identifying and separating  these post-red supergiants from  normal redward evolving yellow
supergiants however are uncertain. 

In Paper I, we suggested that a small group of warm hypergiants with dusty circumstellar 
ejecta were  candidates for post- red supergiant evolution. In this second paper
we have classified several additional late A and F-type supergiants.  
The three most luminous show evidence for circumstellar nebulosity, dusty ejecta, and mass
loss, properties not shared by the other A and F-type supergiants in our sample. 
M33-013442.14 (A8 Ia) has circumstellar dust and strong hydrogen  
P -Cyg profiles (Figures 6 and 11), M31-004424.21 (F5 Ia) has circumstellar nebulosity, and 
M31-004526.62 (A2e Ia) closely resembles the warm hypergiant M31-004444.52 spectroscopically, 
with strong P Cyg profiles in the Balmer and Fe II emission lines plus circumstellar nebulosity, 
although it lacks the 
Ca II and [Ca II] emission and does not have a known  infrared excess. 
We suggest that these stars with evidence for past and current mass loss like the warm hypergiants,
may  also be candidates for post-red supergiant evolution. 

One additional star, M33C-4640, a less luminous A -type supergiant with weak Fe II emission, 
may also be a post-red supergiant candidate, but without any evidence for past or current mass loss.

\subsection{A Very Luminous Star in M33}

M33-013406.63 (B416, UIT 301) is a proposed LBV or LBV candidate \citep{Shemmer,Fabrika}.
Its spectrum shows weak  Fe II and [Fe II] emission, with strong Balmer emission and He I emission. 
The only P Cygni profile is observed in the $\lambda$5876 He I line. It also has  absorption lines of
Si IV, N III, and  He II at $\lambda$4200{\AA}. Consequently, we suggest an O9.5 Ia classification. 

It appears to be  one of the most luminous stars in M33. 
Based on our discussion of its SED (Figure 10c) and interstellar extinction
(Table 5), its M$_{Bol}$ is -12.2 to  -12.7 mag. 
Low amplitude photometric and small spectroscopic variability have been 
reported by \citet{Shemmer} and \citet{Shol04}. On the basis of this 
variability, \citet{Shemmer} suggested that it is an LBV, while based on 
radial velocity variations, \citet{Shol04} propose that it is a B[e] star 
in a close interacting binary with a 16.1 day period. Interestingly, though, the
H$\alpha$ and He I$\lambda$ 5876 emission lines do not vary in phase. 

Given its very high luminosity, it is not surprising that M33-013406.63 may be a
binary or even a multiple star system. Assuming a single star, its luminosity,
$\ge$ 5.9 $\times$ 10$^{6}$ L$_{\odot}$,  
and corresponding spectroscopic temperature, $\approx$ 30,000 K, will place it
far above the LBV/S Dor instability strip on the HR Diagram. 
If we assume two equal mass stars in the system, their luminosities would  
be $\approx$ 3 $\times$ 10$^{6}$ L$_{\odot}$, comparable to $\eta$ Car for 
each. With a higher mass ratio, the primary's luminosity  would exceed $\eta$ Car. 

M33-013406.63 is also near the center of a  ring-shaped bright H II region  in M33. It is tempting to speculate about its possible association with the origin of this nebula. But the H II region's linear dimensions of 104 pc by 81 pc at the distance of M33 plus the lack of the
[N II] $\lambda$ 5755 line \citep{Shemmer} in the H II region spectrum precludes the possibility 
that it is the remnant of a circumstellar nebula  
now being ionized by M33-013406.63  and other hot stars. The star itself, however,  does have a circumstellar nebula \citep{Weis14}.  We conclude that M33-013406.63 is not an LBV/S Dor variable  but a 
very luminous star or pair of stars.

\subsection{Concluding Remarks}

In this first report on our survey of the luminous stars in M31 and M33, we have concentrated
on massive stars with a known or a suspected history of mass loss, and circumstellar ejecta as revealed by their spectra and photometric histories. Based on the above discussion, we have identified three possible  LBV candidates (UIT 008, M31-004425.18, and M33C-15731),  and  some candidates for post-red supergiant evolution. The nature of the Fe II emission line stars, at least those included here, is still uncertain, but many  of them have dusty circumstellar ejecta, which separates them from  the known LBVs. 
 The very luminous late O-type supergiant, M33-013406.63,
is not an LBV, but is one of the most luminous stars or pair of stars, if not the most luminous, in M33. 

Subsequent papers will include additional spectra for a larger set of luminous star 
candidates to be followed by a summary and discussion of the luminous star 
populations in the upper HR diagrams  of  M31 and M33 and other galaxies.

\acknowledgements
Research by R. Humphreys and K. Davidson on massive stars is supported by  
the National Science Foundation AST-1109394. We thank David Thilker for providing a copy of their H I column density map of M31 \citep{Braun}. 
This paper uses data from the MODS1 spectrograph built with funding from NSF grant AST-9987045 and the NSF Telescope System Instrumentation Program (TSIP), with additional funds from the Ohio Board of Regents and the Ohio State University Office of Research.
This publication also makes use of data products from the Wide-field Infrared Survey 
Explorer, which is a joint project of the University of California, Los Angeles, and 
the Jet Propulsion Laboratory/California Institute of Technology, funded by the National Aeronautics and Space Administration.

{\it Facilities:} \facility{MMT/Hectospec, LBT/MODS1}


\begin{deluxetable}{lcccl}
\rotate
\tablewidth{0 pt}
\tabletypesize{\footnotesize}
\tablenum{1} 
\tablecaption{Luminous Stars and Variables in M31 and M33: Spectroscopic Summary}
\tablehead{
\colhead{Star Name} &
\colhead{RA (2000)} &
\colhead{Dec (2000)}  &
\colhead{Spec. Group} &
\colhead{Other Id/Notes/References}  
}
\startdata 
   &    &   M31   &    &    \\
   &  J004229.87\tablenotemark{a} & +410551.8 &  Fe II Em. Line & [Fe II], He I, O I $\lambda$8446 em, see text  \\
   &  J004242.33  & +413922.7 &  Of/late-WN  & P Cyg-type \citet{Massey07}\\ 
   &  J004247.30  &  +414451.0 & Intermed-Type & F2 Ia \\
AE And  &      J004302.52  &  +414912.4 & LBV  & see text, Fig.1  \\
   &  J004313.02   &  +414144.9  & \nodata  & weak-lined, foregrd \\ 
   &  J004318.57   &  +415311.1  & \nodata  & G:V, foregrd  \\ 
   &  J004320.97   &  +414039.6  & Fe II Em. Line  &  O I $\lambda$7774 em , weak [Fe II]\\
   &  J004322.50   &  +413940.9  & Warm Hypergiant  & late A --F0 I (Paper I)  \\
AF And  &   J004333.09 &  +411210.4  & LBV   &  see text, Fig. 5   \\
   &  J004334.50   &  +410951.7  &  Of/late-WN   &  Ofpe/WN9 \citet{Massey07}, nebular em.\\
   &  J004337.16   &  +412151.0  & Intermed-Type & F8 I  \\ 
   &  J004341.84   &  +411112.0  & Of/late-WN    &  P Cyg analog \citet{Massey06b}\\
   &  J004350.50   &  +414611.4  &  Intermed-Type &  A5 I, P Cyg H em. profiles  \\
   &  J004406.32   &  +420131.3  &  Intermed-Type &  F2 Ia  \\ 
   &   J004411.36      &  +413257.2  &  Fe II Em. Line  &  weak Fe II em, He I, nebular em. \\
   &   J004415.00  &   +420156.2     &  Fe II Em. Line  & [Ca II],[Fe II],wk He I em, double H$\alpha$ profile, Fig. 3   \\
   &   J004417.10  &   +411928.0     &  Fe II  Em. Line   &  Ca II, [Ca II], wk He Iem., wk [Fe II], Fig. 3  \\
Var 15  &       J004419.43 &  +412247.0 &  LBV          &  see text           \\
   &   J004424.21  &  +412116.0    &  Intermed-Type & F5 Ia, see text,  Fig. 6  \\
   &   J004425.18\tablenotemark{b}   &  +413452.2   &  Hot Supergiant &  B0-1 Ia, P Cyg H em, broad wings, He I, O I 8446 em, see text  \\
   &   J004442.28  &  +415823.1    &  Fe II Em. Line  & O I $\lambda$7774 em, wk [Fe II], H$\alpha$ asymmetric to blue, two emission features?  \\ 
   &   J004444.52  &   +412804.0   &  Warm Hypergiant & F0 Ia (Paper I) \\
Var A-1  &  J004450.54  &  +413037.7  & LBV          &  see text           \\ 
   &   J004507.65  &  +413740.8    &  Intermed-Type &  A5-A8 I, H em. P Cyg profiles, broad wings \\ 
   &   J004518.76  &  +413630.7    &  Intermed-Type & F2 I  \\
   &   J004522.58  &  +415034.8    &  Warm Hypergiant & A2 Ia (Paper I) \\ 
   &   J004526.62  &   +415006.3   &  Intermed-Type &  A2e I (Fe II em, P Cyg profiles)  see 
   text, Fig. 7 \\ 
   &   J004532.62  &  +413227.8   &  \nodata  & G:V, foregrd  \\
   &   J004535.23  &   +413600.5  &   \nodata & G:V, foregrd \\ 
   &               &              &                               \\
   &               &   M33        &                                \\
Var A & J013232.80 & +303025.0    &  Warm Hypergiant & F8 Ia (Paper I), \citet{RMH87,RMH06} \\
 M33C-4174 &  J013235.25 & +303017.7 & Fe II Em. Line  & He I em. P Cyg, O I $\lambda$7774 em \\
          &   J013242.26 & +302114.1 & Fe II Em. Line  & He I, [Fe II] em., H$\alpha$ asymmetric to blue             \\
UIT 008 &   J013245.41 &  +303858.3 & Of/late-WN & O7-O8f I, see text, (Ofpe/WN9, WR5 \citet{Massey98}), see text \\           
M33C-14239    &  J013248.26 &  +303950.4 & Fe II Em. Line  & He I em.             \\
UIT026    &   J013300.02 &   +303332.4  & \nodata  & underexposed, em lines    \\   
M33C-4640 &  J013303.10 & +303101.9  & Intermed-Type  & A0-A2e Ia, see text, very wk Fe II, wk He I em. \\  
M33C-17194 & J013307.51 & +304258.4  & ----   & composite sp., see text \\  
N025981  &  J013308.95 &  +302956.2  &  Intermed-Type  & A2 Ia, neb. em.,very wk  He I em.  \\ 
M33C-23048 & J013309.12 &  +304954.5 & Hot Supergiant & B0I + WN (Ofpe/WN9, UIT045, WR22 \citet{Massey96}) \\  
M33C-13254 &  J013311.88 & +303853.7  &  Intermed-Type  & B8-A0 Ia $+$ neb em., see text  \\ 
M33C-4119  & J013312.81 & +303012.7  &  Hot Supergiant & B5 Ia, H, He I em., P Cyg \\ 
N033347  & J013316.46 & +303212.0  &  Intermed-Type  & A5-A8 Ia, H$\alpha$ P Cyg    \\ 
          &  J013324.62 &   +302328.4 & Fe II Em. Line  & [Fe II] em., asymmetric H$\alpha$ to blue, double em.       \\ 
M33C-13560 & J013327.26 & +303909.3  &  Of/late-WN & Ofpe/WN9 UIT104, WR39, \citet{Massey96}\\ 
N045901  &  J013327.40  &  +303029.4  & Intermed-Type  & F: pec I, see text, Fig. 6 \\ 
M33C-15742 &  J013332.64 & +304127.2  &  Of/late-WN & WNL WR41 \citet{Massey98} \\ 
M33C-7256 &  J013333.22 &   +303343.4 & Fe II Em. Line  & [Fe II], He I em., neb. em. \\ 
Var C     &      J013335.14 &  +303600.4  &   LBV        &  see text      \\ 
N058746  &  J013335.29 & +304146.0 & foregrd &  G: V + nebular em. \\ 
N061849  &  J013337.00 & +303637.5  &  Intermed-Type  & F5 -F8 Ia + neb. em. \\  
M33C-19725 & J013339.52 & +304540.5  &  Hot Supergiant   & B0.5: I pec\tablenotemark{c}, B517 \citet{HS80}, see text \\  
           &  J013340.6  &  +304137.1  & Hot Supergiant   & O8-O9 I + neb. em. \\  
          &   J013341.28 &    +302237.2   &  Hot Supergiant   &  B2-3  Ia, H, He I em,  P Cyg, 110-A \citet{HS80} \\  
N073136  & J013342.48 &  +303258.5  & Hot Supergiant  &  B8 I + neb em  \\ 
N075866  &  J013343.67 & +303904.0  & \nodata  &  HII region      \\ 
N078046  & J013344.61 & +303559.1   &  Hot Supergiant  & B1-B2 Ia \\  
M33C-18563 &  J013344.78 & +304432.3 &  Hot Supergiant & OB + neb, H,  He I em, UIT187 \citet{Massey96}  \\  
Var B     &     J013349.23 &   +303809.1  &  LBV     &  see text, Fig. 2   \\ 
M33C-15731 &     J013350.12 &  +304126.6 &  Fe II Em. Line & Ca II, [Ca II], [Fe II], wk He I em., H P Cyg,  UIT212 \citet{Massey96}, Fig. 3 \\  
M33C-15235 & J013351.45 &  +304057.0   &  Of/WN  & He I P Cyg profiles, + neb. em.       \\ 
N093351  &     J013352.42 &  +303909.6 & Warm Hypergiant & F0 Ia (Paper I), M33C-13568  \\  
M33C-13206 &  J013353.58 &  +303851.8  &  Of/late-WN & Ofpe/WN9 UIT236, WR103 \citet{Massey98}, asymmetric H profiles \\ 
N097751 &  J013355.17 & +303429.8  &  Hot Supergiant  & early B-type + neb. em.\\ 
B324  &  J013355.96 &  +304530.6 & Warm Hypergiant  &  A8-F0 Ia, UIT 247, (Paper I) \\
       &  J013357.73 &  +301714.2  & Intermed-Type   & A0 Ia, P Cyg profiles \\
M33C-9304  & J013358.70 & +303526.5  &   Hot Supergiant & B0-B1 Ia + WN, H P Cyg, asymmetric to red (B1Ia+WNE, UIT267 \citet{Massey96})  \\ 
N104139  & J013358.92 & +304139.4 &  Intermed-Type  &  F5 Ia, neb. em. \\ 
M33C-8094  & J013359.89 &  +303427.3   &  Hot Supergiant & B0-B1 I + neb. em. \\  
N107775  &  J013401.04 & +303619.5  &  Hot Supergiant & mid-B-type, He II em. very broad + neb em \\ 
M33C-12568  & J013401.91 &  +303819.3  & Of/late-WN & H em profiles very asymmetric to red  \\ 
  &       J013406.63 &   +304147.8 &  Hot Supergiant & O9.5 Ia, see text,B416 \citet{HS80}, UIT301 \citet{Massey96} \\ 
M33C-21386 &  J013406.78 &  +304727.0  &  Of/late-WN & WN7+neb, WR123, UIT 303 \citet{Massey96} \\ 
Var 83  &       J013410.93 &   +303437.6 &  LBV   &  see text        \\ 
N124864  & J013415.19 &  +303704.0  &  Intermed-Type & early A-type + neb em \\ 
N125093  & J013415.38  & +302816.3  &  Warm Hypergiant & F0-F2 Ia, (Paper I) \\ 
M33C-7292 & J013416.10 & +303344.9  & Hot Supergiant & B2.5 Ia, UIT341, B526, two stars, see text \citet{HS80}  \\
        &      J013416.44 &    +303120.8 & Hot Supergiant &  B2-B3 Ia \\  
Var 2   &      J013418.36  &   +303836.9 &  LBV   &  see text, Fig. 5          \\
        &    J013422.91 &  +304411.0  &  Hot Supergiant & B8 Ia, H P Cyg \\  
        &   J013424.78 &    +303306.6 &  Hot Supergiant & B8-A0 Ia, H$\alpha$ asymmetric to blue \\ 
        &   J013426.11  &  +303424.7 & Fe II Em. Line  &  [Ca II], weak [Fe II], He I em, H lines asymmetric to red, Fig. 3  \\
        &   J013429.64 &   +303732.1 &  Hot Supergiant  & B8 Ia, H$\alpha$ P Cyg, broad wings \\ 
        &   J013432.76 &   +304717.2 &  Of/late-WN  & Ofpe/WN9 \citet{Massey07} + strong neb. em.\\  
        &  J013442.14 &   +303216.0  &  Intermed-Type &  A8  Ia, see text, Fig. 6  \\
	&  J013459.47 &   +303701.9  & Fe II Em. Line    & weak [Fe II], He I em.     \\
        &   J013500.30 &  +304150.9  & Fe II Em. Line    &  [Fe II]     \\ 
GR 290   &  J013509.73     &    +304157.3  & Of/late-WN & M33-V532, Romano's star, see text, Fig. 5 \\

\enddata
\tablenotetext{a}{\citet{Massey07} labeled this star peculiar because of an absorption line spectrum characteristic of an F-type star together with Fe II and hydrogen emission lines. This star is only about 1 arcsec from a small cluster in M31 which was probably contaminating its spectrum in their survey. Our spectrum observed with the LBT/MODS1 was obtained under good seeing conditions and the cluster was kept off the  slit. The star is an Fe II emission line star. He I $\lambda$4026 is present in absorption, and there are no F-type absorption lines.}   
\tablenotetext{b}{The spectrum may have changed since \citet{Massey07}. The blue spectrum shows 
absorption lines typical of  early B-type supergiants while a strong O I $\lambda$7774 line common in A to F-type supergiants is present in the red.}
\tablenotetext{c}{Several of the He I absorption lines appear to be double 
including $\lambda$4026,4387,4471, and 6678 due to emission in the line core. H$\alpha$ and H$\beta$ are asymmetric with  a second emission component on the red side.}
\end{deluxetable}

\begin{deluxetable}{llllllllllllllllll}
\rotate
\tablewidth{0 pt}
\tabletypesize{\scriptsize}
\tablenum{2} 
\tablecaption{Multi-Wavelength Photometry }
\tablehead{
\colhead{Star} & 
\colhead{U}  &
\colhead{B} &
\colhead{V} &
\colhead{R} & 
\colhead{I} &
\colhead{J} &
\colhead{H} &
\colhead{K} &
\colhead{3.6$\mu$m}\tablenotemark{a} &
\colhead{4.5$\mu$m}\tablenotemark{a} &
\colhead{5.8$\mu$m}\tablenotemark{a} &
\colhead{8$\mu$m}\tablenotemark{a}  &
\colhead{3.4$\mu$m}\tablenotemark{b} &
\colhead{4.6$\mu$m}\tablenotemark{b} &
\colhead{12$\mu$m}\tablenotemark{b} &
\colhead{22$\mu$m}\tablenotemark{b} & 
\colhead{Var}
}

\startdata
   &    &   &    &  & & & & M31   & &   &  &  &  &   & &   \\ 
M31-004229.87 & 18.5&	19.1	&18.8 &	18	& 18.2 &  \nodata & \nodata & \nodata & 12.4&	11.6&	11	&10.2&		12.6	&11.7	&9.5	&  7.2 &  \\  
M31-004242.33 &  17.85	&18.7	&18.6	& 18.2	& 18.1  & \nodata & \nodata & \nodata &  \nodata & \nodata & \nodata & \nodata & \nodata & \nodata & \nodata & \nodata & \\ 
M31-004247.30 & 17.1&	16.9	&16.4	&16	& 15.6 & 15.3	& 15	& 15 &  \nodata & \nodata & \nodata & \nodata & 14.55 & 	14.44&	11.78&	8.66&  \\  
AE And  & 16.35	& 17.25	& 17.4	& 17.2	& 17.2 & 16.39& 15.89&	15.8 &  15.84	&15.43	&14.93	& 12.84	& 	15.4	& 15.1	& 11.1	& 7& LBV   \\  
M31-004320.97 &  19	& 19.8	& 19.2  & 	18.4	& 17.9 & 17.1&	16.7	& 15.2 &  13.3	& 12.7	& 11.8	& 10.9	& 	13.5	& 12.5	& 9.2	& 7.2&   \\ 
M31-004322.50  & 20.7& 	21.2	& 20.3	& 19.9	& 19.2 & \nodata & \nodata & \nodata & 14.9& 	14.6& 	14.4  & 13.5 & 15.7&	15.5&	12.3&	9.1&  \\   
AF And &  16.4& 	17.3	& 17.3	&17.5	& 17.4 & 15.8	& 15.37 & 	15.41 & \nodata & \nodata & \nodata & \nodata &  14.8&	15&	11.3& 	8& LBV  \\  
M31-004334.50 & 17.3	& 18.2	& 18.1& 	17.8& 	17.7 &  16.5	& 15.6	& 14.7 &  \nodata & \nodata & \nodata &  \nodata & \nodata & \nodata & \nodata & \nodata&   \\ 
M31-004337.16 &     19.6	&18.9	&17	&16.55	&16.15 & 16.5& 	15.6	& 14.7  & \nodata & \nodata & \nodata &  \nodata & 14.95&	14.11&	11.06&	9.27&  \\    
M31-004341.84 & 17.2& 	17.95&	17.5 & 	17.1	& 16.8    & 16.4	& 16.2	& 15.6  & 15.6	& 14.8	& 14.2	& 13.1 & 
\nodata & \nodata & \nodata &  \nodata&  \\  
\enddata
\tablenotetext{a}{{\it Spitzer}/IRAC}
\tablenotetext{b}{WISE}
\end{deluxetable}

\begin{deluxetable}{llcccl}
\rotate
\tablewidth{0 pt}
\tabletypesize{\footnotesize}
\tablenum{3} 
\tablecaption{Stars with Circumstellar Nebulae and Dust }
\tablehead{
\colhead{Star Name} &
\colhead{Spec. Group} &
\colhead{[NII]} &
\colhead{Ca II/[CaII]} & 
\colhead{IR excess} & 
\colhead{Comments}  
}
\startdata 
   &    & M31   &   &   &    \\
M31-004229.87  &  Fe II em. line &  \nodata & \nodata & yes & CS dust \\ 
M31-004242.33  &  Of/late-WN &   yes & \nodata & \nodata & CS nebula?\\  
AE And         &   LBV     & yes  & \nodata & yes & CS nebula, f-f em., H II PAH \\          
M31-004320.97 & Fe II Em. Line   & yes & \nodata & yes & CS nebula, CS dust \\
M31-004322.50 & Warm Hypergiant   &  \nodata &  yes  & yes & CS dust \\
AF And  & LBV   & yes  & \nodata  & yes  & CS nebula, f-f em., H II PAH  \\
M31-004341.84  & Of/late-WN  & yes & \nodata & yes  & CS nebula, f-f em., CS dust(?)\\
M31-004411.36  &  Fe II Em. Line  &  yes &   \nodata &  yes & CS nebula, H II PAH \\
M31-004415.00  &  Fe II Em. Line  &  \nodata &  yes & yes & CS dust \\
M31-004417.10  &  Fe II  Em. Line   &  yes &  yes & yes & CS dust, CS nebula \\
Var 15  &  LBV &  yes & \nodata &  yes & CS nebula, H II PAH   \\
M31-004424.21 & Intermed-Type F5 Ia & yes & \nodata & yes & CS nebula?, f-f em, H II PAH\\ 
M31-004442.28  &  Fe II Em. Line   &  \nodata &  \nodata & yes & CS dust \\
M31-004444.52  &  Warm Hypergiant  &  \nodata &  yes & yes & CS dust \\
Var A-1  & LBV  & yes & \nodata&  \nodata &   CS nebula \\ 
M31-004522.58  &  Warm Hypergiant &  yes  & yes & yes & CS dust\tablenotemark{a}, CS nebula \\ 
M31-004526.62 &  Intermed-Type  A2e Ia  & yes & \nodata & \nodata & CS nebula, f-f em. \\ 
   &      &         &     &         &    \\
   &     &   M33   &    &      &      \\
Var A  &  Warm Hypergiant & \nodata &  yes & yes & CS dust \\
M33C-4174  & Fe II Em. Line   &  yes & \nodata & yes & CS nebula, HII PAH \\
M33-013242.26 &  Fe II Em. Line  &  yes  & \nodata & yes & CS nebula, H II PAH \\
UIT 008 &   Of/late-WN  O7-O8f I & yes & \nodata & yes & H II PAH\\ 
M33-013248.26  & Fe II Em. Line  & yes  & \nodata & yes & H II PAH \\
M33-013324.62 & Fe II Em. Line   & \nodata & \nodata & yes & CS dust  \\ 
N045901  &  F I: pec &  weak  & \nodata &  yes & f-f em:?., CS nebula?, H II PAH \\ 
M33C-7256 &   Fe II Em. Line  & yes & \nodata & yes & CS nebula, CS dust \\ 
Var C   &   LBV   & yes & \nodata & \nodata   &  CS nebula \\ 
Var B     &  LBV   & yes & \nodata & yes & CS nebula, H II PAH \\ 
M33C-15731 & Fe II Em. Line  & yes & yes & yes & CS nebula, CS dust\\  
N093351\tablenotemark{b}  & Warm Hypergiant F0 Ia  & \nodata & yes & yes & CS dust \\  
B324  &  Warm Hypergiant  A8-F0 Ia  & yes & yes & yes & f-f em., H II PAH \\
M33-013406.63  &  Hot Supergiant  O9.5 Ia & yes & \nodata & yes & CS nebula, H II PAH \\ 
Var 83  &  LBV   & yes & \nodata & \nodata & CS nebula, f-f em.  \\ 
N125093 &  Warm Hypergiant  F0-F2 Ia  & \nodata & yes & yes & CS dust\\ 
Var 2   &  LBV    & yes & \nodata & \nodata & CS nebula  \\
M33-013426.11 & Fe II Em. Line  & \nodata & yes & yes & CS dust \\
M33-013442.14 &  Intermed-Type  A8 Ia  & \nodata & \nodata & yes & CS dust  \\
M33-013459.47 & Fe II  Em. Line    & yes & \nodata & yes & CS nebula, f-f em. + CS dust(?), H II PAH\\  
M33-013500.30 &  Fe II  Em. Line    & yes  & \nodata & yes & CS nebula, CS dust\\ 
GR 290/V532 & Of/late-WN &   yes & \nodata & \nodata & CS nebula \\
\enddata
\tablenotetext{a}{In Paper I we noted that the IRAC photometry was discrepant and suggestive of H II
contamaination. The IRAC source is more than 5$\arcsec$ away and in H II nebulosity while the WISE source agrees with the position of the star.}
\tablenotetext{b}{N093351 is a good example of the role of nebular contamination in the spectra from nearby emission nebulosity. The MMT/Hectospec spectrum shows strong nebular emission lines in the 
1\farcs5 aperture while the same lines are not present in the LBT/MODS1 spectrum with a 0.6\arcsec slit and sky subtraction immediately on either side of the star.} 
\end{deluxetable}

\begin{deluxetable}{lccccccc}
\tablewidth{0 pt}
\tabletypesize{\scriptsize}
\tablenum{4} 
\tablecaption{Extinction and Total Luminosities }
\tablehead{
\colhead{Star} &
\colhead{A$_{v}$ ((H I)} &
\colhead{A$_{v}$ (H I(Seizert)} &
\colhead{A$_{v}$ (colors)} &
\colhead{A$_{v}$ (stars)}  &
\colhead{Adopted A$_{v}$} & 
\colhead{M$_{v}$} &
\colhead{M$_{Bol}$} \\
    &   (mag)  & (mag)  &  (mag)  & (mag) & (mag) & (mag) & (mag) 
}
\startdata 
    &        &  M31       &     &   &  &  & \\
AE And (LBV) &   0.9      &  0.6   &   \nodata & 0.9 (1) &  0.6  & -7.6   & -9.3 to -9.8  \\
AF And (LBV) &   1.1      &  1.0   &  \nodata &  \nodata & 1.0   &  -8.1  & -10.7  \\
Var A-1 (LBV) &  1.9       &  \nodata & \nodata & 1.5 (1) &  1.5    & -8.8   & -10.1:  \\
Var 15 (LBV)  &  1.3       &  \nodata & \nodata & 0.3 (1)  & 1.3  &  -7.3  &  -8.5:  \\
M31-004229.87 (Fe II em) & 2.1  &  \nodata & \nodata & 1.4 (2) &  1.4 & -7.0  & -8.1:  \\
M31-004320.97 (Fe II em) & 1.5  &  \nodata & \nodata & \nodata & 1.5     & -6.7  &  -7.8:  \\
M31-004411.31 (Fe II em) & 0.8   & \nodata & \nodata & 2.3 (3)  &  2.3 & -8.6  & -9.7:   \\ 
M31-004415.00 (Fe II em) & 0.9  & \nodata & \nodata & 2.7 (2) & 2.7 & -8.8  &  -9.6:  \\
M31-004417.10 (Fe II em) & 1.7  & \nodata & \nodata & 0.7 (1) & 0.7 & -8.0&  --8.3:  \\
M31-004442.28 (Fe II em) & 1.7   & \nodata & \nodata & 1.2 (1) & 1.2 & -5.9  & -8.3:  \\ 
M31-004526.62 (A2e Ia)   & 1.3  & \nodata & 1.5     & 1.6 (1) & 1.5  & -8.7 &  -8.9 \\
M31-004424.21  (F5 Ia)   &  1.9 & \nodata &  1.9     & 1.6 (4) & 1.9  & -9.6 &  -9.5  \\
M31-004425.18 (B2-B3 I)  &  0.7 & \nodata &  \nodata & \nodata & 0.7  & -7.7 &  -8.0 \\  
   &        &         &     &   &  &  &  \\
    &        &   M33      &     &   &  &  &  \\
Var B (LBV)  & 0.3(min.)   & 0.7 &  \nodata & \nodata &  0.7   & \nodata   &  -10.4  \\ 
Var C (LBV)  & 0.4     & 0.85 &  \nodata &   0.7 (1) &  0.85  & -8.1  &  -9.8 \\
Var 83 (LBV) & 0.9     & 1.05 &  \nodata &  0.85 (2) &  1.0   & -9.5   & -11.1 \\  
Var 2 (LBV) & 1.0     & \nodata & \nodata & 0.2 (1)  & 1.0 &  -7.3     &  -9.1: \\   
M33C-4174 (Fe II em) & 0.9  & \nodata  & \nodata & 0.8 (3) & 0.8  & -7.3    &  -8.7:  \\
M33-013242.26 (Fe II em) & 0.8  &  \nodata  & \nodata & 0.9 (2) & 0.9 & -8.0   &  -8.9:\\
M33-013248.26 (Fe II em) & 0.4 & \nodata  & \nodata  & 0.3 (2) & 0.4 & -7.6  &  -8.9: \\
M33-013324.62 (Fe II em) & 0.4  & \nodata  & \nodata  & \nodata & 0.4 & -5.3  &  -6.3:\\
M33C-7256 (Fe II em)     &   0.3(min)      &  \nodata  & \nodata & 0.5 (3) & 0.5  &  -5.6 & 7.6:\\
M33C-15731 (Fe II em)    &  0.7  & \nodata  & \nodata & 1.2 (2) & 1.2  & -8.9   &  -10.9 \\
M33-013426.11 (Fe II em) & 0.6   & \nodata  & \nodata & \nodata  & 0.6     & -6.1  &  -6.9:\\
M33-013459.47 (Fe II em) & 0.7 & \nodata  & \nodata & \nodata  & 0.7      & -6.8  &  -7.5:\\
M33-013500.30 (Fe II em) & 0.7 & \nodata  & \nodata &  0.3 (3)  & 0.7 & -5.5 & -6.9: \\ 
UIT 008 (O7-O8f/WN9)  &  0.8   & \nodata  &   0.6   &   0.4(1)  & 0.6  &  -7.5  & -10.7 \\ 
M33-013406.63 (O9.5 I)   & 0.7(1.2) & \nodata  & 1.5 & 0.3(2), 0.8(1)  & 0.7(1.2) & -9.1(-9.6) & -12.2(-12.7)\\
M33C-4640 (A0-A2e Ia)   & 0.6      & \nodata  &  0.4   &  0.65(2)  &  0.6   &  -8.1  &  -8.3  \\ 
M33-013442.14 (A8 Ia)    & 1.2  & \nodata  &  2.1  &  0.45 (1) & 2.1 & -9.3  &  -9.2\\
M33-N045901 ( F I pec)   & 0.5  & \nodata  &  1.6  &  0.9 (1)  & 1.6 & -8.5  & -8.4\\
Gr290/V532  (Of/WN)      & 0.6  &  \nodata  & \nodata & \nodata & 0.6\tablenotemark{a} &  var. & -10.4 - -10.7\\ 
\enddata
\tablenotetext{a}{Romano's star. In addition to the estimated A$_{v}$ from the H I, two nearby associations, OB 88 and  OB 89 yield A$_{v}$ $\approx$ 0.5 mag. The star is variable. We adopt the magnitudes at visual maximum and minimum and the corresponding temperatures from \citet{Shol} together with the bolometric corrections using the B.C discussed by \citet{Pol2011} for visual maximum. This yields a range for M$_{bol}$ of -10.4 to -10.7 at visual maximum and -10.5 at minimum.} 
\end{deluxetable}

\begin{deluxetable}{lcccc}
\tablewidth{0 pt}
\tabletypesize{\footnotesize}
\tablenum{5} 
\tablecaption{Outflow Velocities (km s$^{-1}$) }
\tablehead{
\colhead{Star} &
\colhead{P Cyg (H)} &
\colhead{P Cyg (He I)} &
\colhead{P Cyg (Fe II)} & 
\colhead{Double H$\alpha$}
}
\startdata
      &       &   LBVs   &      &     \\
AE And(B2-3)  &  -151(2)$\pm$ 2.0  &  -161(2)$\pm$ 4.0  &  \nodata  &  \nodata  \\
AF And(Of/WN)  &  -245(3)$\pm$ 4.5  &  -222(4)$\pm$ 4.6  &  -213([Fe III])(2)$\pm$ 2.0  &  \nodata \\
Var 15(Of/WN)  &  -219(4)$\pm$ 3.1  &   -205(6)$\pm$ 10 &  \nodata &  \nodata  \\
Var A-1        &  -200(4)$\pm$ 2.8  &  -160(2)$\pm$ 5.0  &  -219(1)  &   \nodata  \\
Var C(B1-2, quiescence)    &  -158(4)$\pm$ 8  &  \nodata  &  -157(3)$\pm$ 3  &   \nodata  \\
Var B(B0-B1)   &  -230(2)$\pm$ 3.0  &  -221(3)$\pm$ 2.7   &  \nodata  &  \nodata   \\
Var 83         &  -158(4)$\pm$ 6.8  &  \nodata   &  -147(3)$\pm$ 3.0  &  \nodata   \\
Var 2(Of/WN)   &  -221(1)  &  -236(4)$\pm$ 4.1   &  \nodata  &  \nodata    \\ 
      &       &      &      &     \\
      &       &  Fe II Em. Line  &       &      \\
M31-004415.00 &  \nodata &   \nodata &  \nodata  &  -120,+155 \\
M31-004417.10 &  \nodata &  -269(1)  &   \nodata &  \nodata  \\
M33C-4174     &  \nodata &  -168(1)  &  \nodata  &   \nodata \\
M33C-15731    &  -216(3)$\pm$ 8.0 &  \nodata  &  \nodata  &   \nodata  \\
M33-013426.11 &  -242(2)$\pm$ 5.7 &  \nodata  &  \nodata  &   \nodata  \\ 
      &       &      &      &     \\
      &       &   Warm Hypergiants\tablenotemark{a}  &      &     \\ 
M31-004322.50(late A-F0)\tablenotemark{b}  &  -240(1)  &  \nodata &  -185(3)$\pm$ 3 & \nodata  \\
M31-004444.52(F0 Ia)\tablenotemark{b} &  -244(3)$\pm$ 2 &  \nodata &  -214(3)$\pm$ 9 &  \nodata     \\
M31-004522.58(A2 Ia) &  \nodata &  \nodata & \nodata & $\pm$ 100 \\
N093351(F0 Ia) &  -127(1)  &  \nodata &  \nodata &   $\pm$ 120 \\
B324(A8-F0 Ia)\tablenotemark{b}  &  -143(1)  &  \nodata &  \nodata &  \nodata  \\ 
N125093(F0-F2 Ia) &  -262(1) & \nodata &  \nodata & $\pm$ 220  \\ 
      &       &                           &        &        \\
      &       &   Other Stars (Table 4)   &        &        \\ 
GR 290/V532(Of/WN - 2010)  &   \nodata &  -234(4)$\pm$ 5.0 &  \nodata &     \\ 
M33-013406.63(O9.5 Ia) & \nodata  &  -307(1)  &  \nodata &   \nodata \\
M31-004526.62(A2e Ia)  &  -146(4)$\pm$ 4.6 &  \nodata  &  -139(3)$\pm$ 2.5 &   \nodata \\ 
M33-013442.14(A8 Ia)   &  -221(2)$\pm$ 8.2        &           &          &           \\ 
UIT 008(O7-O8/WN9) &      -159(1) &  -138(1) & \nodata &   \nodata \\
M31-004425.18(B2-B3 I) &  -153(3)$\pm$ 8.9 & \nodata  &  \nodata &  \nodata \\
N045901(F: I) &  \nodata  &  \nodata &  \nodata & $\pm$ 90 \\
      &       &      &      &     \\
            &       &   Normal Supergiants   &        &        \\
Of/late-WNs(9 stars) &  -329(7)$\pm$ 23 &  -313(20)$\pm$ 15.6 & \nodata &   \nodata \\
Early B(B0-B3)(12 stars) & -220(15)$\pm$ 5& -213(9)$\pm$ 9 & \nodata &   \nodata \\
Late B(B5-B8)(13 stars) & -161(19)$\pm$ 7& -160(3)i$\pm$ 3.4&  \nodata &  \nodata  \\
A type(A0-A8)(6 stars) & -142(9)$\pm$ 5.4 &  \nodata &  \nodata &  \nodata  \\  
\enddata
\tablenotetext{a}{The velocities for the warm hypergiants (Paper I) are repeated here for completeness and due to wavelength calibration problems with the LBT spectra which have been corrected here.}
\tablenotetext{b}{P Cyg velocities for the Ca II[Ca II] profiles, M31-004322.50: -90, M31-004444.52: -200, B324: -126.} 
\end{deluxetable}

\begin{deluxetable}{lc}
\tablewidth{0 pt}
\tabletypesize{\footnotesize}
\tablenum{6} 
\tablecaption{Mass Lost Estimates}
\tablehead{
\colhead{Star} &  
\colhead{Mass Lost (M$_{\odot}$)} IR (SED)   
}
\startdata 
M31-004229.87 (Fe II em.)  & 1.82 $\times$ 10$^{-2}$      \\
M31-004320.97 (Fe II em.)          & 0.95 $\times$ 10$^{-2}$  \\
M31-004415.00 (Fe II em.)     & 1.66  $\times$ 10$^{-2}$     \\ 
M31-004417.10  (Fe II em.)    &  0.38  $\times$ 10$^{-2}$     \\
M31-004442.28  (Fe II em.)  & 0.22  $\times$ 10$^{-2}$     \\  
M33-013324.62  (Fe II em.)   &  0.92 $\times$ 10$^{-2}$  \\
M33C-7256   (Fe II em.)   & 0.56  $\times$ 10$^{-2}$    \\ 
M33C-15731   (Fe II em.)  & 1.40   $\times$ 10$^{-2}$    \\
M33-013426.11 (Fe II em.) &  0.32 $\times$ 10$^{-2}$    \\
M33-013442.14 (A8 Ia)    & 0.56  $\times$ 10$^{-2}$    \\
M33-013459.47  (Fe II em.)    &  0.29 $\times$ 10$^{-2}$    \\
M33-013500.30  (Fe II em.)   &  0.42 $\times$ 10$^{-2}$    \\
\enddata
\end{deluxetable}

\begin{deluxetable}{llllllllllllllllll}
\rotate
\tablewidth{0 pt}
\tabletypesize{\scriptsize}
\tablenum{2-on line} 
\tablecaption{Multi-Wavelength Photometry }
\tablehead{
\colhead{Star} & 
\colhead{U}  &
\colhead{B} &
\colhead{V} &
\colhead{R} & 
\colhead{I} &
\colhead{J} &
\colhead{H} &
\colhead{K} &
\colhead{3.6$\mu$m}\tablenotemark{a} &
\colhead{4.5$\mu$m}\tablenotemark{a} &
\colhead{5.8$\mu$m}\tablenotemark{a} &
\colhead{8$\mu$m}\tablenotemark{a}  &
\colhead{3.4$\mu$m}\tablenotemark{b} &
\colhead{4.6$\mu$m}\tablenotemark{b} &
\colhead{12$\mu$m}\tablenotemark{b} &
\colhead{22$\mu$m}\tablenotemark{b} &  
\colhead{Var}
}

\startdata
   &    &   &    &  & & & & M31   & &   &  &  &  &   & &   \\ 
M31-004229.87 & 18.5&	19.1	&18.8 &	18	& 18.2 &  \nodata & \nodata & \nodata & 12.4&	11.6&	11	&10.2&		12.6	&11.7	&9.5	&  7.2  \\  
M31-004242.33 &  17.85	&18.7	&18.6	& 18.2	& 18.1  & \nodata & \nodata & \nodata &  \nodata & \nodata & \nodata & \nodata & \nodata & \nodata & \nodata & \nodata \\ 
M31-004247.30 & 17.1&	16.9	&16.4	&16	& 15.6 & 15.3	& 15	& 15 &  \nodata & \nodata & \nodata & \nodata & 14.55 & 	14.44&	11.78&	8.66 \\  
AE And  & 16.35	& 17.25	& 17.4	& 17.2	& 17.2 & 16.39& 	15.89&	15.8 &  15.84	&15.43	&14.93	& 12.84	& 	15.4	& 15.1	& 11.1	& 7 &  LBV\\  
M31-004320.97 &  19	& 19.8	& 19.2  & 	18.4	& 17.9 & 17.1&	16.7	& 15.2 &  13.3	& 12.7	& 11.8	& 10.9	& 	13.5	& 12.5	& 9.2	& 7.2  \\ 
M31-004322.50  & 20.7& 	21.2	& 20.3	& 19.9	& 19.2 & \nodata & \nodata & \nodata & 14.9& 	14.6& 	14.4  & 13.5 & 15.7&	15.5&	12.3&	9.1 \\   
AF And &  16.4& 	17.3	& 17.3	&17.5	& 17.4 & 15.8	& 15.37 & 	15.41 & \nodata & \nodata & \nodata & \nodata &  14.8&	15&	11.3& 	8 & LBV  \\  
M31-004334.50 & 17.3	& 18.2	& 18.1& 	17.8& 	17.7 &  16.5	& 15.6	& 14.7 &  \nodata & \nodata & \nodata &  \nodata & \nodata & \nodata & \nodata & \nodata  \\ 
M31-004337.16 &     19.6	&18.9	&17	&16.55	&16.15 & 16.5& 	15.6	& 14.7  & \nodata & \nodata & \nodata &  \nodata & 14.95&	14.11&	11.06&	9.27 \\    
M31-004341.84 & 17.2& 	17.95&	17.5 & 	17.1	& 16.8    & 16.4	& 16.2	& 15.6  & 15.6	& 14.8	& 14.2	& 13.1 & 
\nodata & \nodata & \nodata &  \nodata \\  
M31-004350.50 &  18	& 18.3	& 17.7	& 17.2	& 16.75  & 16.25& 	15.8& 	15.9  & \nodata & \nodata & \nodata &  \nodata &   \nodata & \nodata & \nodata &  \nodata & var?\\ 
M31-004406.32 &    &     &    &    &     &  14.8& 	14.7	& 14.4  &  \nodata & \nodata & \nodata &  \nodata & \nodata & \nodata & \nodata & \nodata  \\
M31-004411.36 &  18.3	& 18.8	&  18.1 & 	17.5	& 17  &  16.6	& 16.4	& 15.5  &  14.9	& 14.5	& 13.3	& 11.7 & \nodata & \nodata & \nodata &  \nodata & var  \\  
M31-004415.00 &  17.8	& 18.6	& 18.3	& 17.3	& 17.2  & 16.8	& 16	& 14.5  & 12.5	&  11.7	& 11.7	& 10.3	&13	& 12&	10	& 8.4  \\   
M31-004417.10 &  16.5& 	17.2	& 17.1 & 	16.8	& 16.6  & 15.9	& 15.6	& 14.7  & 14.9	& 14	& 12.9	& 11.9 & \nodata & \nodata & \nodata &  \nodata \\
Var 15 &  17.9	& 18.5	& 18.4  & 17.0 	& 17.9	 & 16.0 &	15.6&	15.3 &   14.9	&14.6&	14.2&	12.9& 		15.3&	14.9	& 11.4	& 7.2 &  LBV \\   
M31-004424.21 &  18.4	& 17.6	& 16.7	& 16.2  & \nodata &  15	& 14.8	& 14.6 &  \nodata & \nodata & \nodata &  \nodata & 13.53	& 13.09 & 	8.67& 	6.35 & var?\\ 
M31-004425.18 & 16.7	&17.6	& 17.5	& 17.3	& 17.15  &  16.7	& 16.15	& 16.9 &  \nodata & \nodata & \nodata &  \nodata & \nodata & \nodata & \nodata & \nodata  &  var?\\ 
M31-004442.28 & 18.6& 	19.8& 	19.7& 	19& 	19.1 &  \nodata & \nodata & \nodata & 14.7&	14	&13.4	& 12.5&		14.9	& 14.1	& 10.7	& 9.1  \\  
M31-004444.52  &  19	& 19.1	& 18.1	& 17.3	& 16.6 & 15.8&	15.2&	14.4 & 12.6	& 11.8	& 11.2	& 10.2 & 12.6	&11.7	&8.6	& 6.5 &  var \\   
Var A-1 & 16.7  & 17.35 & 17.1  & 16.8  & 16.6  & 15.75 &  15.54        & 15.46  &  \nodata & \nodata & \nodata &  \nodata & \nodata & \nodata & \nodata & \nodata & LBV  \\
M31-004507.65   &   15.9	& 16.4	& 16.15	& 15.9 &  \nodata & 15.4& 	15.3& 	15  & \nodata & \nodata & \nodata &  \nodata & \nodata & \nodata & \nodata & \nodata  \\
M31-004518.76  &  17.4	&17.2	&16.7	&16.7	& 16.4 &  15.4	& 15.1	& 15  & \nodata & \nodata & \nodata &  \nodata & 14.46 &14.17&	10.94&	9.38 \\  
M31-004522.58 & 17.9&	18.6& 	18.5& 	18.2	& 18 & 17& 	16.3& 	15.4 & 15.4	& 15.6	&  13.7	&  11.9  & 14.2	&13.3	&10.8	&8.5 \\  
M31-004526.62 &  16.75	&17.7	& 17.2	& 16.8	& 16.7  & 15.9	& 15.7	& 15.1  &  \nodata & \nodata & \nodata &  \nodata & \nodata & \nodata & \nodata & \nodata  \\ 
   &               &              &                               \\
      &    &   &    &  & & & & M33   & &   &  &  &  &   & &   \\
M33-Var A\tablenotemark{c}& 20.2: & 19.9  & 19.1  & 18.6 & \nodata  & 16.9  & 15.9 &  14.7 & 12.9   & 12.1 & 11.4  &  10.1  & 13.30& 	12.10&	8.80	& 7.40 & Paper I \\
M33C-4174 & 17.10&	18.05	&18.0	&17.9	& 17.7 &  \nodata & \nodata & \nodata & 16.3	15.9 &  \nodata & \nodata & 15.5&	15.2	&11.3	& 9.0  & \nodata \\   
M31-013242.26 & 17.0&	18.1&	17.4&	16.55 & \nodata & 15.1	& 14.2& 	13.85 & 13.7&	13.6	&\nodata 	13.2 	&	13.5 &	13.5 & 	11.2 & 	8.9  & \nodata \\   
UIT 008  &  15.9 &	17.5&	17.6&	17.6& 	17.6  &  \nodata & \nodata & \nodata & 15.3& \nodata & \nodata & \nodata &
14.9&	14.2	&9.4&	6.1 & \nodata \\  
M33-013248.26  &   16.1 &	17.3&	17.25	&17.0&	16.9&  \nodata  &\nodata & \nodata & 15.7&	15.4&	15.0&	14.6&			14.9 &	14.6&	13.1&	9.5  & \nodata \\   
UIT026  &  17.1	& 18.2	& 18.3	& 18.4	& 18.4  & \nodata & \nodata & \nodata & 15.9	& 16.2  & \nodata & \nodata & \nodata & \nodata & \nodata & \nodata  & \nodata \\ 
M33C-4640 &  16.5&	17.2	&17	&16.8	&16.8  & 16.7&	16.1&	16.15  &  16.2	& 16.5  \nodata & \nodata & \nodata & \nodata & \nodata &  \nodata & not var.\\ 
N025981  &  & & & & &  16.9 &	15.9&	15.45  &  \nodata & \nodata & \nodata &  \nodata & 14.3	&13.7&	8.7&	6.2 & \nodata  \\ 
M33C-23048 &  16.8 &	17.9&	17.9&	17.8&	17.8  & \nodata & \nodata & \nodata & 17.3&	17.0 & \nodata & \nodata & \nodata & \nodata & \nodata & \nodata & \nodata \\
M33C-13254 &  16.2&	16.9 &	16.8& 	16.7& 	16.7  &  14.7&	14.5&	14.4 & 14.3	& 14.3&  \nodata & \nodata & 		13.4&	12.9&	8.1&	5.1 &  var.? \\   
M33C-4119 &  16.7&	17.5	&17.3	&17	& 17 & \nodata & \nodata & \nodata & 15.8&	14.9& \nodata &	13.1 &  \nodata & \nodata & \nodata &  \nodata & var.? \\  
N033347 & 16.7 &	17.1&	16.85&	17.05&	16.8  &  16.0&	16.0 & 	15.8  & 15.65	& 15.8  & \nodata & \nodata & \nodata & \nodata & \nodata & \nodata  & \nodata \\
M33-013324.62  &  18.65 &	19.5 &	19.6& 	19.3& 	19.5  &  \nodata &	\nodata&	\nodata  & 14.3	& 13.6&  13.0 & 12.3 & 		14.4&	13.4&	11.2&	9.2 & \nodata   \\   
M33C-13560 & 16.75	&17.8&	17.95&	17.9&	17.9 &  \nodata & \nodata & \nodata &  \nodata & \nodata & \nodata &  \nodata &   \nodata & \nodata & \nodata &  \nodata & var.? \\ 
N045901\tablenotemark{d}  &  18.5&	18.3&	17.6	&17.2	& 16.8 &  16.4: & 15.0: & 14.8: &  \nodata & \nodata & \nodata &  \nodata &   13.6 & 12.9 & 10.3 &  8.0   & \nodata  \\ 
M33C-15742 & 18.1	&18.9	&19	&18.8	&18.7 &  \nodata & \nodata & \nodata &  \nodata & \nodata & \nodata &  \nodata &   \nodata & \nodata & \nodata &  \nodata & \nodata \\
M33C-7256 &  18.3	&19.3&	19.4&	18.4&	18.9 & 17.1 &	15.9&	14.8  &  13.1&	12.5&	12.2 &	11.6&	13.4&	12.5&	9.4&	7.3  & \nodata  \\   
Var C\tablenotemark{e} &    15.5 & 	16.5&	16.4&	16.3&	16.1   & 16.6&	16.25&	15.7 &  14.5&	14.4&	14.5&	
14.2 &  \nodata & \nodata & \nodata &  \nodata & LBV \\ 
N061849 &  17.8 & 	17.3&	16.55	&16.2	&15.8  & 15.3&	15.1   &15.0  & 16.6 & \nodata & \nodata &  \nodata &   \nodata & \nodata & \nodata &  \nodata & \nodata  \\
M33C-19725 & 16.6&	17.6&	17.5&	17.4&	17.3  &  16.8&	17.15&	15.3   & \nodata & \nodata & \nodata &  \nodata  &  \nodata & \nodata & \nodata &  \nodata & var.\\
M33-013340.6  & \nodata 	&19.9	&18.3	&18.1	& 18.2  & \nodata & \nodata & \nodata &  \nodata & \nodata & \nodata &  \nodata &   \nodata & \nodata & \nodata &  \nodata & \nodata \\
M33-013341.28  &   15.2  & 	16.2& 	16.3&	16.3&	16.3  & 16.3	&15.7&	15.4  & 15.1&	15.0&	15.1&	14.5&	
15.4&	15.1&	11.6&	9.0 & \nodata   \\    
N073136 & 17.2&	18.1&	17.6&	17.0&	16.3  &  15.1&	14.4&	14.05  &  15.6	&15.6& \nodata & \nodata &	13.5&	13.3&	10.8&	6.6 & \nodata \\  
N078046  &   16.2&	17.2&	17.2&	17.2&	17.2  & \nodata & \nodata & \nodata & 13.30	&13.30	& \nodata & 12.10
& 13.2&	13.0&	11.1&	8.9 & \nodata  \\  
M33C-18563 & 17.75&	18.5&	18.2&	18	&17.8  &  \nodata & \nodata & \nodata &  \nodata & \nodata & \nodata &  \nodata &   \nodata & \nodata & \nodata &  \nodata & \nodata  \\
Var B &  15.3&	16.2&	16.2&	16	& 15.9   & 14.25&	14.07&	13.95  &  15.7	& 15.9   & \nodata & \nodata & \nodata &  \nodata & \nodata & \nodata & LBV \\ 
M33C-15731 &  15.8&	16.9&	16.8&	16.4&	16.3  &  15.4	&14.7	&14.05  & 12.2&	11.7&	11.1&	10.6&			12.3&	11.6i&	9.2&	7.6 & not var.  \\    
M33C-15235 & 16.9&	17.8&	17.7&	17.6&	17.5  &  \nodata & \nodata & \nodata &  \nodata & \nodata & \nodata &  \nodata &   \nodata & \nodata & \nodata &  \nodata & var.? \\
N093351 &  16& 	16.4& 	16.2&	16& 	15.7 & 15.7&	15.3& 	14.7 & 13.0 & 12.6 & \nodata & 10.3 & \nodata & \nodata & \nodata & \nodata &  Paper I \\ 
M33C-13206 &  17.15&	18.0&	18.15&	18.1&	18.1  &  \nodata & \nodata & \nodata &  \nodata & \nodata & \nodata &  
\nodata & 12.5 &	13.2&	9.1&	6.5  & \nodata \\   
N097751  &  16.7&	17.6	&17.6&	17.6&	17.5  &  15.9&	15.0&	14.7 &  \nodata & \nodata & \nodata &  \nodata & 
14.4&	14.4&	10.7&	8.9  & \nodata  \\   
B324\tablenotemark{f} & 14.9 & 15.3	& 14.9	& 15.2	& 15.0 & 13.7& 	13.4& 	13.3 & 12.7&	12.5  & \nodata & \nodata &  12.50&	12.15&  	8.30& 	5.90 & Paper I \\  
M33-013357.73 & 18.8 & 17.4&	17.4&	17.3&	17.2  &  \nodata & \nodata & \nodata & 16.7&	16.5& \nodata & \nodata &
16.3&	15.9&	13.0&	9.3 & \nodata \\    
M33C-9304  &  15.8&	16.9	&16.7	&16.6	& 16.6   &  16.4&	15.95&	15.6  &  15.2	&14.6 & \nodata & 12.1 & 
\nodata & \nodata & \nodata &  \nodata & not var. \\
N104139 &  18.3 & 	17.8&	17.2&	16.5 &  \nodata & 15.6&	15.4&	15.3  &  14.8&	14.8& \nodata & \nodata & 		14.2&	14.2&	9.6&	7.5 & \nodata  \\   
M33C-8094  &  16.9	&17.8&	17.9	&17.9	& 18.1  & \nodata & \nodata & \nodata & 14.9&	15.1 &  \nodata & \nodata & \nodata &  \nodata & \nodata & \nodata &  var.? \\ 
N107775  &  17.2&	18.1&	18.1&	18.0 &	17.8  &  16.7	&16.0&	15.1  & 15.3&	15.9  & \nodata & \nodata & \nodata &  \nodata & \nodata & \nodata & \nodata \\
M33C-12568 &   16.9&  	17.8	&17.9	&17.7	& 17.8  &  \nodata & \nodata & \nodata & \nodata & \nodata & \nodata &  \nodata &   \nodata & \nodata & \nodata &  \nodata & not var. \\
M33-013406.63  &  15.1&	16.3&	16.1&	15.9&	15.8  &  15.4	&15.4&	15.0  & 14.4&	14.0&	13.8& \nodata & 	
14.0&	13.6&	8.4&	6.2 &  see text  \\   
M33C-21386 &  16.5	&17.3	&17.4	&17.4	& 17.3  &  \nodata & \nodata & \nodata & \nodata & \nodata & \nodata &  \nodata &   \nodata & \nodata & \nodata &  \nodata & \nodata  \\
Var 83 & 15.1	&16.1	&16.0	&15.9	&15.7  &  15.6&	15.4  & 15.3 & 15.0&	14.8 &\nodata & \nodata & \nodata & \nodata & \nodata & \nodata &  LBV \\ 
N124864  & 19.4 &	18.65&	17.6&	17.1&	16.5  &  15.9&	15.65	&15.45  &  15.3& 	15.3 &  \nodata & \nodata & \nodata & \nodata & \nodata & \nodata & \nodata \\
N125093 & 18.4&	18.1&	17.3& 	16.8 &	16.2 & 15.4&	14.8&	14.1 & 12.5& 	11.9&  \nodata  & 10.0 & 12.90	& 11.90	 & 8.90	& 6.40 & \nodata  \\  
M33-013416.44 &  16.1	&17.1	&17.1	&17.0  &17.0  &  \nodata & \nodata & \nodata & 15.3&	15.2  &  \nodata & \nodata & \nodata & \nodata & \nodata & \nodata & \\
Var 2 &  17.0&	18.1&	18.2&	18.1	&18.0 &  \nodata & \nodata & \nodata &  16.3&	16.3 & \nodata & \nodata & \nodata & \nodata & \nodata & \nodata & LBV \\
M33-013422.91 &  16.35	&17.3&	17.2&	17.1&	17.0  & \nodata & \nodata & \nodata & 16.4&	16.6  & \nodata & \nodata & \nodata & \nodata & \nodata & \nodata & \nodata  \\
M33-013424.78 &  16.2&	17.0&	16.8&	16.7&	16.5   &  16.3&	16.8&	16.3  & 15.2&	15.0&	14.7&	14.0&			15.7&	15.5&	10.7&	7.6 & \nodata  \\     
M33-013426.11  &  18.4	&19.2&	19.0&	18.6&	18.3   &  \nodata & \nodata & \nodata & 14.6&	13.8&	13.1i&	12.2& 		15.0&	13.8&	10.8&	8.9 & \nodata  \\   
M33-013429.64  &   16.2&	17.1&	17.1&	17.0&	16.9  &  16.8&	16.05	&15.8  &  \nodata & \nodata & \nodata & \nodata & 	15.0&	13.8&	11.8&	9.0 & \nodata \\  
M33-013432.76 &   17.6	&18.8	&19.1	&19.0 &	18.8   &  \nodata & \nodata & \nodata &  \nodata & \nodata & \nodata &  \nodata &   \nodata & \nodata & \nodata &  \nodata & \nodata \\
M33-013442.14 &   18.4&	18.2&	17.3&	16.9&	16.4  & 16.0&	15.2&	14.6  & 13.6&	13.0&	12.4&	11.6&			13.6&	12.9&	10.3&	8.0  & \nodata \\    
M33-013459.47 &  17.3&	18.6&	18.4&	17.9&	17.7   &  \nodata & \nodata & \nodata & 14.4&	13.7&	13.2&	12.3&		14.6&	13.7&	11.8&	8.7 & \nodata \\    
M33-013500.30 & 18.4&	19.2&	19.3&	18.6&	19.1  &  \nodata & \nodata & \nodata & 13.9&	13.2&	12.5&	11.9&		14.0&	13.0&	11.0&	8.9 & \nodata  \\   
GR 290 & \nodata       & \nodata   & var.   & \nodata    & \nodata  &   16.8&	16.7&	16.8  &  16.3	&15.9&  \nodata & \nodata & \nodata &  \nodata & \nodata & \nodata &  see text\\ 		
\enddata
\tablenotetext{a}{{\it Spitzer}/IRAC}
\tablenotetext{b}{WISE}
\tablenotetext{c}{The photometry is from \citet{RMH06}. Recent CCD photometry
obtained at the Barber Observatory, Univeristy of Illinois Springfield, in October and December 2012 suggests that 
Var A may have begun to brighten.  Its current  V magnitude is 18.36 $\pm$ 0.05 
 measured relative to more than 40 comparison stars in the same field. }
\tablenotetext{d}{The two closest sources in the 2MASS catalog are offset from the target by 
$\approx$ 3\arcsec at
01:33:27.26 +30:30:32.1 and 01:33:27.29 +30:30:32.4 and with nearly identical JHK magnitudes. We suspect that
they may be the same source. The identification is therefore doubtful. The closest IRAC source is at 01:33:27.81 +30:30:28.2.}  
\tablenotetext{e}{The LBVs are variable, and both Var C and Var B experienced LBV-type eruptions during the past decade. The 2MASS, IRAC, and WISE photometry was not observed at the same
 time as the visual phoometry from \citet{Massey07}.  The visual magnitudes for Var C, however,  are 
 nearly the same 
 as the B and V magnitudes published by \citet{Szeif}.  We therefore adopted the corresponding JHK 
 photometry from \citet{Szeif} for its SED in Figure 8. } 
\tablenotetext{f}{B324 is in a very crowded region and groundbased visual 
photometry of B324 can be  contaminated by nearby faint stars within an arcsec
or so of B324. For that reason we measured magnitudes from HST/WFPC2 F555W and F439W obtained November 30, 1998. Converted to the standard V and B magnitudes give 15.12 and 15.56 respctively. Slightly fainter than the groundbased photometry. }
\end{deluxetable}

\end{document}